\documentclass[]{bytedance_seed}

\usepackage[toc,page,header]{appendix}

\usepackage{minitoc}

\usepackage{url}

\usepackage{soul}
\usepackage{ulem}
\usepackage{wrapfig}

\usepackage{minted}
\usepackage{xcolor}

\usepackage[utf8]{inputenc}
\usepackage{xcolor}
\usepackage{algorithm}
\usepackage{algpseudocode}

\usepackage{booktabs}   
\usepackage{multirow}   
\usepackage{makecell}   
\usepackage{array}      

\usepackage{amsmath}    
\usepackage{amsfonts}   

\usepackage{tcolorbox}
\tcbuselibrary{skins, breakable}

\def\ours{\textsc{UniEP}} 

\title{\ours{}: Unified Expert-Parallel MoE MegaKernel for LLM Training}

\author[1,2]{Size Zheng}
\author[1]{Xuegui Zheng}
\author[1]{Li-wen Chang}
\author[2]{Jidong Zhai}

\affiliation[1]{ByteDance Seed}
\affiliation[2]{Tsinghua University}

\abstract{
The exponential growth in Large Language Model (LLM) parameters has transformed model training into an increasingly resource-intensive endeavor. With the stagnation of Moore's Law and the widening disparity between computation throughput and communication bandwidth, expert parallelism (EP) has emerged as a critical strategy for scaling mixture-of-experts (MoE) models. However, despite numerous proposals for optimizing EP, ranging from communication compression to computation-communication overlap, adoption within production-grade frameworks like Megatron-LM remains conservative. Existing solutions often rely on ad-hoc, complex kernels that lack adaptability across diverse optimization configurations and frequently neglect numerical stability, failing to meet the strict precision requirements of large-scale training.

In this paper, we introduce \ours{}, a novel system that unifies diverse EP optimization strategies into a cohesive abstraction. \ours{} fuses the MoE communication and computation into {MegaKernels}, effectively transforming complex architectural tuning into a unified parameter search space for automated adaptability. Crucially, \ours{} incorporates a deterministic token ordering mechanism that guarantees numerical consistency with sequential execution, even under aggressive overlap schedules. We evaluate \ours{} on GPU clusters equipped with NVIDIA Hopper GPUs. Our results demonstrate that \ours{} achieves 1.03$\times$-1.38$\times$ speedups over state-of-the-art work, effectively mitigating communication bottlenecks while maintaining the rigorous accuracy standards required for production LLM training.
}

\date{\today}

\begin{document}
\maketitle


\section{Introduction}

The landscape of deep learning has been fundamentally altered by the emergence of mixture-of-experts (MoE) architectures, which increases model size at scalable training and inference cost. State-of-the-art large language models (LLMs), including GPT-5~\cite{gpt5}, Gemini3 Pro~\cite{gemini3pro}, DeepSeek-V3~\cite{deepseek-v3}, and Qwen3~\cite{qwen3} have universally adopted MoE designs to scale parameter counts into hundreds of billions while maintaining manageable activation budgets. This architectural paradigm extends beyond text, underpinning the latest multimodal models such as Qwen-VL~\cite{qwen-vl} and DeepSeek-VL~\cite{deepseek-vl}. As the number of experts per layer grows (typically ranging from 64 to 384), expert parallelism (EP) has become the de facto standard for distributing these models across GPU clusters. In this regime, GPUs exchange tokens to route them to their designated experts, a process that introduces significant communication overhead. Empirical profiling suggests that the MoE-specific components (combining dispatch, expert computation, and combine phases) can consume $30\%$ to $80\%$ of the total training budget~\cite{netmoe, comet, moescale}, making the optimization of EP a critical determinant of overall training efficiency.

\textbf{Motivation.}
The primary challenge in scaling EP lies in the growing disparity between GPU computation throughput and interconnect bandwidth. While GPU tensor performance has skyrocketed, interconnect bandwidth (e.g., NVLink/InfiniBand) scales at a significantly slower pace. Consequently, EP training is frequently bound by the All-to-All or All-Gather communication phases required for token routing. Although numerous optimization techniques~\cite{comet, deepep, megablocks, moeblaze, hybrid-ep, hetermoe} exist, their integration into production environments faces strict constraints regarding numerical correctness and hardware efficiency. The objective of this paper is to bridge the gap between communication-computation overlap techniques and the rigorous requirements of production-grade training, specifically addressing the inefficiencies inherent in current separate-kernel execution models.

\textbf{Limitation of state-of-the-art approaches.}
The community has proposed diverse strategies to mitigate EP bottlenecks, including dynamic load balancing (e.g., MegaBlocks~\cite{megablocks}), model compression (DeepSpeed-MoE~\cite{deepspeed-moe}), heterogeneous offloading (HeterMoE~\cite{hetermoe}), and communication-computation overlap (MegaScale-MoE~\cite{moescale}, COMET~\cite{comet}). Specialized communication libraries like DeepEP~\cite{deepep} have also been developed to accelerate communication primitives. Production frameworks such as Megatron-LM~\cite{megatron-lm} now integrate these advancements, combining Transformer Engine~\cite{te} with optimized kernels~\cite{deepep, megablocks} to push performance limits.

Despite these strides, we identify two critical deficiencies in state-of-the-art work.
First, existing approaches treat computation and communication as distinct, isolated operations, relying on coarse-grained pipelining (e.g., CUDA streams) to achieve overlap. This separation necessitates frequent host-side intervention for stream synchronization. In distributed settings, rank synchronization often leads to bubbles that negate the benefits of overlap. More critically, coarse-grained overlap often requires splitting micro-batches, which introduces numerical instability. Due to the non-associativity of floating-point arithmetic (especially in BFloat16), splitting the backward pass alters the accumulation order of gradients, rendering the training process sensitive to system-level scheduling and compromising reproducibility.
Second, current frameworks lack a unified communication primitive. The optimal strategy fluctuates based on workload characteristics: AllGather is bandwidth-efficient when the number of routed tokens (top-$k$) is large~\cite{moescale}, whereas AllToAll is superior for sparse routing~\cite{deepep}. This dichotomy forces developers to maintain complex heuristics and separate kernels, complicating performance tuning and portability.

\textbf{Key insights and contributions.}
To resolve these limitations, we introduce \ours{}, a unified system that optimizes EP training through fine-grained MegaKernel fusion. Our approach is driven by two key insights.

First, we observe that optimal overlap can be achieved without host-side multi-stream management by leveraging the GPU's streaming multiprocessors (SMs) for block-level scheduling. We propose applying {MegaKernel} techniques, which is previously restricted to inference contexts~\cite{mirage, hazymega}, to the training domain. Unlike inference-focused approaches that aggressively fuse the entire LLM into one kernel, we strategically fuse only the MoE-specific sub-graphs (Dispatch+GroupGEMM and GroupGEMM+Combine). Within our MegaKernel, distinct SMs are dynamically assigned to computation or communication roles. Dependencies are managed via an on-chip scoreboard mechanism at the token granularity. This design allows for aggressive, fine-grained overlap within a single CUDA stream, eliminating CPU overheads and synchronization bubbles while maintaining high SM occupancy. Crucially, by avoiding micro-batch splitting, we preserve the deterministic order of operations, guaranteeing numerical consistency with sequential execution.

Second, we unify the disparate communication patterns through a parameterized abstraction. \ours{} can achieve lower communication volume of both AllGather and AllToAll without explicit kernel switching. We abstract the optimization space of this MegaKernel (e.g., tile sizes, SM allocation, warp allocation) into a unified parameter set. Instead of manual tuning, we provide a rigorous hardware performance model that automatically identifies optimal configurations for varying input shapes and cluster topologies.

Our specific contributions are as follows:
\begin{itemize}
    \item We present the first application of MegaKernel architecture to MoE training, enabling high-performance communication-computation overlap within one stream while strictly preserving numerical precision.
    \item We introduce a unified EP abstraction utilizing token remapping and parameter space search, which automates the adaptation to diverse MoE workloads and eliminates the complexity of manual primitive selection.
    \item We evaluate \ours{} on production-scale workloads (including DeepSeek~\cite{deepseek-v2, deepseek-v3, deepseek-vl}, Qwen~\cite{qwen3, qwen-vl} variants), and Kimi~\cite{kimi-k2} across clusters equipped with NVIDIA Hopper GPUs. Experiments demonstrate significant speedups over 
    baselines in both kernel-level benchmarks and end-to-end training on up to 128 GPUs.
\end{itemize}

\textbf{Experimental methodology and artifact availability.}
Our evaluation is conducted on modern clusters equipped with NVIDIA Hopper GPUs. We analyze performance across two types of Hopper GPUs with varying compute-to-bandwidth ratios and diverse configurations of micro-batch sizes and sequence lengths. We benchmark \ours{} against the highly optimized MoE modules that incorporate current SOTA kernels (including DeepEP~\cite{deepep}, Transformer Engine~\cite{te}, and COMET~\cite{comet}). To validate real-world applicability, we deploy \ours{} in full-scale training. The MegaKernel implementation has been made open-source (\url{https://github.com/ByteDance-Seed/Triton-distributed}) to facilitate further research.

\section{Background}
\label{sec:background}

This section details the architectural workflow of EP training and explain the techniques of state-of-the-art work.

\begin{figure}[!t]
    \centering
    \includegraphics[width=\textwidth]{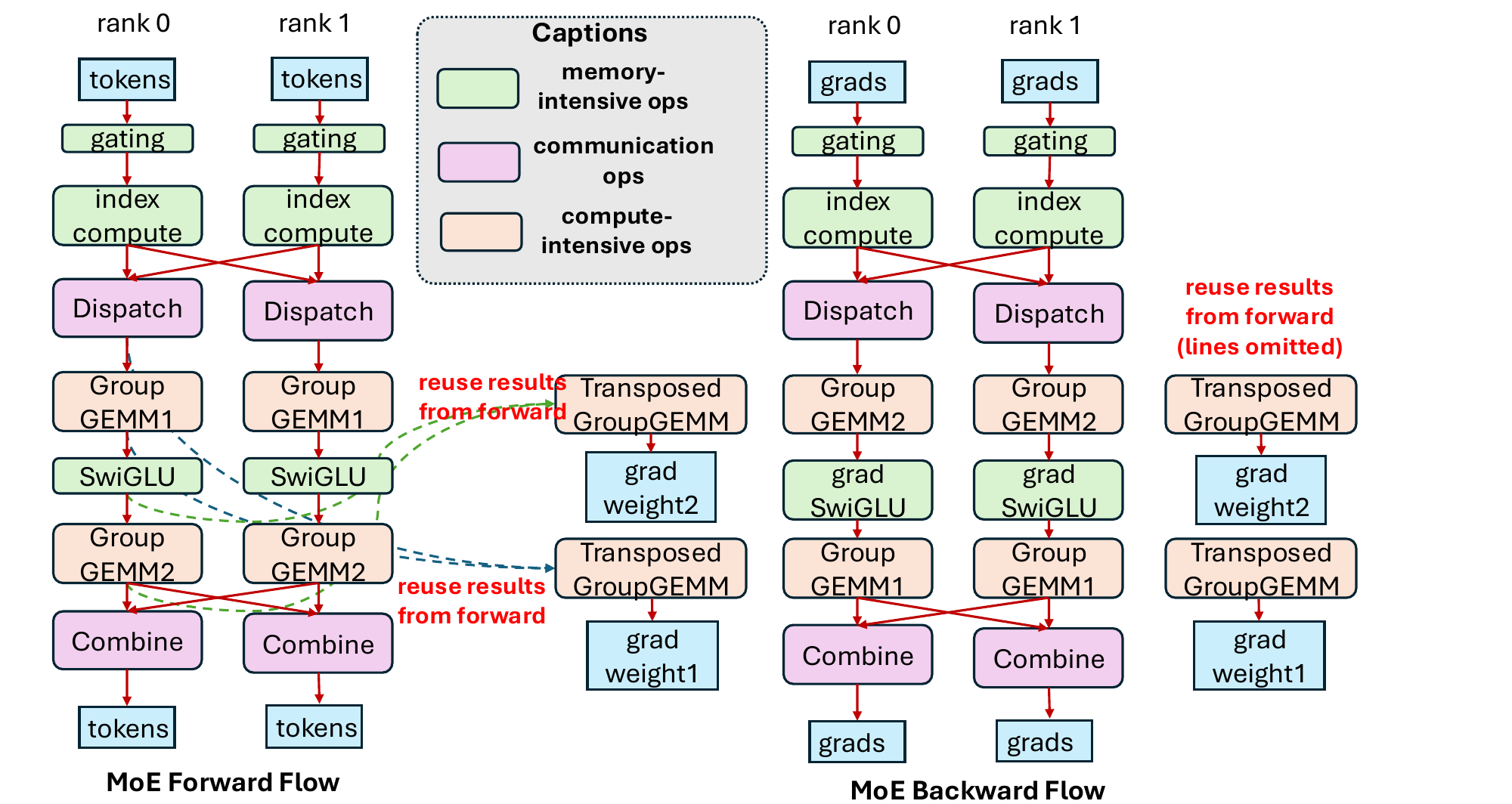}
    \caption{MoE forward and backward flow.}
    \label{fig:moe_structure}
\end{figure}

\subsection{MoE Training Workflow}
The workflow of an MoE layer involves complex computation and communication, as illustrated in Figure~\ref{fig:moe_structure}.

\textbf{Forward Propagation.}
The process initiates with a {gating} (or routing) phase, where a router network calculates the probability distribution for each token across available experts. Based on these probabilities, the system determines the destination rank and expert ID for each token. A key parameter is top-$k$ (typically $k=8$), defining how many experts process each token.
Once routing decisions are finalized, tokens must be permuted to group them by destination expert. While the destination rank is mathematically determined, the intra-expert {token ordering} (by index compute) depends on the implementation. To ensure reproducibility, frameworks typically enforce a deterministic order (e.g., stable sort preserving original sequence, or a fixed permutation). However, inconsistencies in sorting algorithms across different kernel implementations can lead to varying execution orders. After sorting, the {Dispatch} phase transmits tokens to their target ranks. This is implemented via an \texttt{AlltoAll} primitive or an \texttt{AllGather} followed by a local scatter.
Upon arrival, tokens undergo the compute phase: a GroupGEMM operation projects inputs to the hidden dimension, followed by a SwiGLU activation, and a second GroupGEMM projects them back. Finally, the {Combine} phase aggregates results. Tokens are routed back to their source ranks (via \texttt{AlltoAll} or \texttt{ReduceScatter}) and weighted by their gating scores.

\textbf{Backward Propagation and Numerical Stability.}
The backward pass mirrors the forward sequence but introduces significant complexity regarding numerical precision. Gradients with respect to the output are dispatched to experts to compute gradients for inputs and weights. The calculation of weight gradients necessitates a {Transposed GroupGEMM}. Unlike the forward pass GroupGEMM, which partitions data along the spatial (token count) dimension, the Transposed GroupGEMM partitions along the accumulation dimension (corresponding to the batch and sequence length axes).
This distinction is critical for system design. If an optimization strategy (e.g., pipeline overlap) splits the input batch into micro-batches to hide latency, it inherently alters the accumulation tree of the Transposed GroupGEMM. This causes bit-wise divergence in gradients compared to sequential execution. Consequently, aggressive overlap strategies often compromise strict numerical equivalence to the baseline.

\subsection{State-of-the-Art Optimizations}
Recent research has focused on three pillars of MoE efficiency: computation, communication, and overlap.

\noindent
\textbf{Computation: Padding-Free GroupGEMM.}
The number of tokens assigned to each expert is dynamic and workload-dependent. Naive implementations use padding to uniform shapes, wasting compute cycles. State-of-the-art systems like MegaBlocks~\cite{megablocks} and vLLM~\cite{vllm} implement padding-free GroupGEMMs based on Block-sparse formats. These kernels process tokens in fixed-size tiles (e.g., $128 \times 256$) and utilize dynamic pointer arithmetic to handle irregular offsets. By eliminating padding, these kernels significantly reduce memory pressure and achieve high hardware utilization, usually exceeding 65\% of the theoretical peak FLOPS on NVIDIA Hopper architectures. This tiled execution model also serves as the prerequisite for fine-grained pipelining.

\noindent
\textbf{Communication: NVSHMEM-based Primitives.}
To overcome the latency of standard collective libraries (NCCL~\cite{nccl}), systems like COMET~\cite{comet} and DeepEP~\cite{deepep} leverage NVSHMEM~\cite{nvshmem} for direct peer-to-peer memory access. COMET optimizes the \texttt{AllGather} primitive, while DeepEP provides a specialized \texttt{AlltoAll} implementation. A common trade-off in these libraries is the reservation of compute resources for communication; for instance, DeepEP may dedicate a subset of SMs (e.g., 20) exclusively to drive network traffic, thereby reducing the throughput available for the main compute kernel.

\noindent
\textbf{Overlap Strategies.}
Hiding communication latency requires overlapping Dispatch/Combine phases with GroupGEMM computation.
{Tile-level Overlap:} This approach decomposes the workload into tiles. Communication and computation are launched in separate CUDA streams. Synchronization is managed via fine-grained signals in global memory; once the \texttt{AllGather} stream completes a tile, it flags the compute stream to proceed.
COMET uses this overlap approach.
{Batch-level Overlap:} This method employs double buffering at the micro-batch level. While stream 1 processes the computation of batch $i$, stream 2 executes the communication for batch $i+1$. DeepEP uses such approach.
A pervasive limitation in both approaches is the reliance on multi-stream execution. Managing dependencies across streams necessitates CPU intervention for kernel launching and synchronization, introducing host-side overheads. Furthermore, as noted previously, batch-level splitting strategies often sacrifice bit-wise reproducibility for performance.
\section{\ours{} Overlapping MegaKernel Design}
\label{sec:design}

To address the limitations of coarse-grained overlapping approaches (e.g., dual-stream or micro-batch splitting), we propose \ours{}, a unified execution model that fuses the entire computation and communication graph into a single {MegaKernel}. Unlike prior works~\cite{comet, flux} that rely on the host CPU to manage dependency chains, \ours{} offloads the entire scheduling logic to the GPU SMs. This design enables fine-grained, instruction-level overlap between dispatch, computation, and combine phases without compromising numerical precision.

In this Section, we detail the architectural design of the MegaKernel, focusing on the worker abstraction, the deterministic token mapping algorithm, and the dynamic scheduling mechanism.
As illustrated in Figure~\ref{fig:moe_structure}, the pattern of {Dispatch} followed by {GroupGEMM} and GroupGEMM followed by Combine constitutes the primary bottleneck in the MoE forward pass. We fuse these operations into a cohesive kernel to maximize overlap opportunities while maintaining modularity.

\subsection{Dispatch+GroupGEMM MegaKernel Design}
\label{sec:dispatch_groupgemm}

\textbf{Persistent Worker Architecture}
Our MegaKernel adopts a persistent threadblock approach. Upon kernel launch, we launch a fixed number of threadblocks exactly equal to the number of physical SMs on the GPU. Each thread block functions as a persistent {Worker}, occupying the SM for the kernel's duration. This one-to-one mapping eliminates the overhead of context switching and prevents deadlocks that could arise from inter-block synchronization if the number of blocks exceeded hardware capacity.

Within the MegaKernel, a worker can dynamically assume one of three roles:
\begin{itemize}
    \item \emph{Communication Worker (Comm-Worker):} Responsible for sending tokens and issuing inter-GPU transactions.
    \item \emph{Computation Worker (Comp-Worker):} Responsible for executing GroupGEMM tiles.
    \item \emph{Relay Worker:} Manages synchronization signals and performs intra-GPU data multicasting.
\end{itemize}

\textbf{Deterministic Token Mapping}
A critical requirement for \ours{} is to maintain bitwise numerical equivalence with standard sequential execution. Since the {Comm-Worker} must transmit tokens based on index computation, we implement a global addressing scheme that translates local token layouts to a deterministic global order.

The mapping process constructs a translation table from a local token ID to a tuple: (target\_rank, expert\_id, destination\_offset). To ensure that the final order of tokens in the destination expert's buffer is deterministic regardless of parallel execution order, we propose the mapping strategy outlined in Algorithm~\ref{alg:token_mapping}.

\begin{algorithm}[t]
\caption{Deterministic Global Token Mapping}
\label{alg:token_mapping}
\begin{algorithmic}[1]
\Require $E_{sel}$: Selected experts per token $[N_{tok}, topk]$
\Require $M_{loc}$: Local stable sorting indices $[N_{tok}, topk]$
\Ensure $M_{glob}$: Global mapping $[N_{tok}, topk] \to (R, E, O)$

\State $C_{exp} \gets \text{BinCount}(E_{sel}, \text{num\_bins}=N_{exp})$
\State $O_{exp} \gets \text{CumSum}(C_{exp}, \text{dim}=-1)$
\State $C_{all} \gets \text{AllGather}(C_{exp})$ \Comment{Shape: $[W, N_{exp}]$}

\State \Comment{Calculate global offset for each expert per rank}
\State $O_{all} \gets \text{ComputeGlobalOffsets}(C_{all})$

\For{$r \gets 0$ \textbf{to} $W-1$} \Comment{Iterate over ranks}
    \For{$t \gets 0$ \textbf{to} $N_{tok}-1$}
        \For{$j \gets 0$ \textbf{to} $topk-1$}
            \State $e_{id} \gets E_{sel}[t, j]$
            \State $r_{tgt} \gets e_{id} // N_{epr}$
            \State $e_{loc} \gets e_{id} \% N_{epr}$
            
            \State $\text{base\_off} \gets O_{all}[r_{tgt}, e_{loc}, r]$
            \State $\text{loc\_idx} \gets M_{loc}[t, j]$
            \State $\text{final\_idx} \gets \text{loc\_idx} - O_{exp}[e_{id}] + \text{base\_off}$
            
            \State $M_{glob}[t, j] \gets (r_{tgt}, e_{loc}, \text{final\_idx})$
        \EndFor
    \EndFor
\EndFor
\end{algorithmic}
\end{algorithm}

This algorithm accepts any local stable sorting permutation ($M_{loc}$) and extends it to a global context using an AllGathered offset table ($O_{all}$). 
The function \textsc{ComputeGlobalOffsets} (Line 4) is critical for ensuring deterministic placement. It resolves the exact memory address offset for every token in the global address space without requiring atomic locking during runtime.
Specifically, let $C_{all}[i, e]$ denote the number of tokens that Rank $i$ intends to send to Expert $e$. This matrix is obtained via the \texttt{AllGather} collective. Since tokens destined for Expert $e$ are stored contiguously in the destination buffer, ordered by their source rank (Rank $0 \rightarrow \dots \rightarrow \text{Rank } W-1$), the write offset for a token from Rank $r$ to Expert $e$ is determined by the total count of tokens sent to $e$ by all preceding ranks.
Mathematically, the global offset base $O_{all}[r, e]$ is computed as the exclusive prefix sum along the rank dimension:
\begin{equation}
    O_{all}[r, e] = \sum_{k=0}^{r-1} C_{all}[k, e]
\end{equation}
By combining this base global offset with the local stable sort index (Line 13), each Comm-Worker can independently calculate the unique, conflict-free destination address for every token. This mechanism allows massive parallelism in the communication phase while strictly preserving the token order required for numerical reproducibility.
By strictly adhering to this pre-calculated map, the {Comm-Worker} guarantees that tokens arrive in memory in a fixed order, preserving the summation order in subsequent reduction phases. The Comm-Worker utilizes vectorized loads (128-bit) per thread to maximize memory bandwidth, with the degree of parallelism tunable by adjusting the number of active warps.

\textbf{Scoreboard-based Synchronization}
To coordinate the producers (Comm/Relay) and consumers (Comp), we implement a lightweight synchronization mechanism via a {global memory scoreboard}. The scoreboard is a structured array in global memory, partitioned into two segments:
\begin{equation}
    \text{Scoreboard} = [\; \underbrace{\mathcal{S}_{token}}_{\text{Token Arrival}} \;|\; \underbrace{\mathcal{S}_{tile}}_{\text{Tile Ready}} \;]
\end{equation}
The $\mathcal{S}_{token}$ segment tracks the arrival of individual tokens, while $\mathcal{S}_{tile}$ signals the readiness of a complete GroupGEMM tile. 
Different workers use the scoreboard differently:

\begin{itemize}

\item  \emph{Relay Worker Logic:} The Relay Worker monitors incoming tokens. When a Comm-Worker successfully writes a token to the destination buffer, it updates the corresponding entry in $\mathcal{S}_{token}$. The Relay Worker polls this entry; upon confirmation, it increments an internal tile counter. Once the counter matches the GroupGEMM tile size (e.g., 128), the Relay Worker atomically sets the flag in $\mathcal{S}_{tile}$.

\item  \emph{Comp Worker Logic:} The Comp Worker polls $\mathcal{S}_{tile}$. Upon receiving the ready signal, it executes the computation for the specific tile. We employ a padding-free GroupGEMM kernel (inspired by MegaBlocks~\cite{megablocks} and vLLM~\cite{vllm}) that uses dynamic pointer arithmetic to handle irregular tile shapes, ensuring high Tensor Core utilization.
    
\end{itemize}

\textbf{Dynamic Scheduling}
Static assignment of SM roles often leads to load imbalance, particularly given the variance in MoE routing loads. To address this, \ours{} employs a dynamic role reassignment strategy.
We maintain a global atomic counter in device memory that serves as a task queue cursor. The task space is linearized: indices $[0, N_{comm})$ correspond to communication tasks, while $[N_{comm}, N_{comm}+N_{comp})$ correspond to computation tasks.

At runtime, an SM acquires a task by atomically incrementing the global counter. The returned ID determines its immediate role:
\begin{enumerate}

\item  If the ID maps to a communication task, the SM becomes a Comm-Worker.
\item   If the ID maps to a computation task, the SM becomes a Comp-Worker (potentially waiting on the scoreboard).
\item   When a task is completed, the SM requests a new ID.
    
\end{enumerate}

\begin{figure}[!t]
    \centering
    \includegraphics[width=\textwidth]{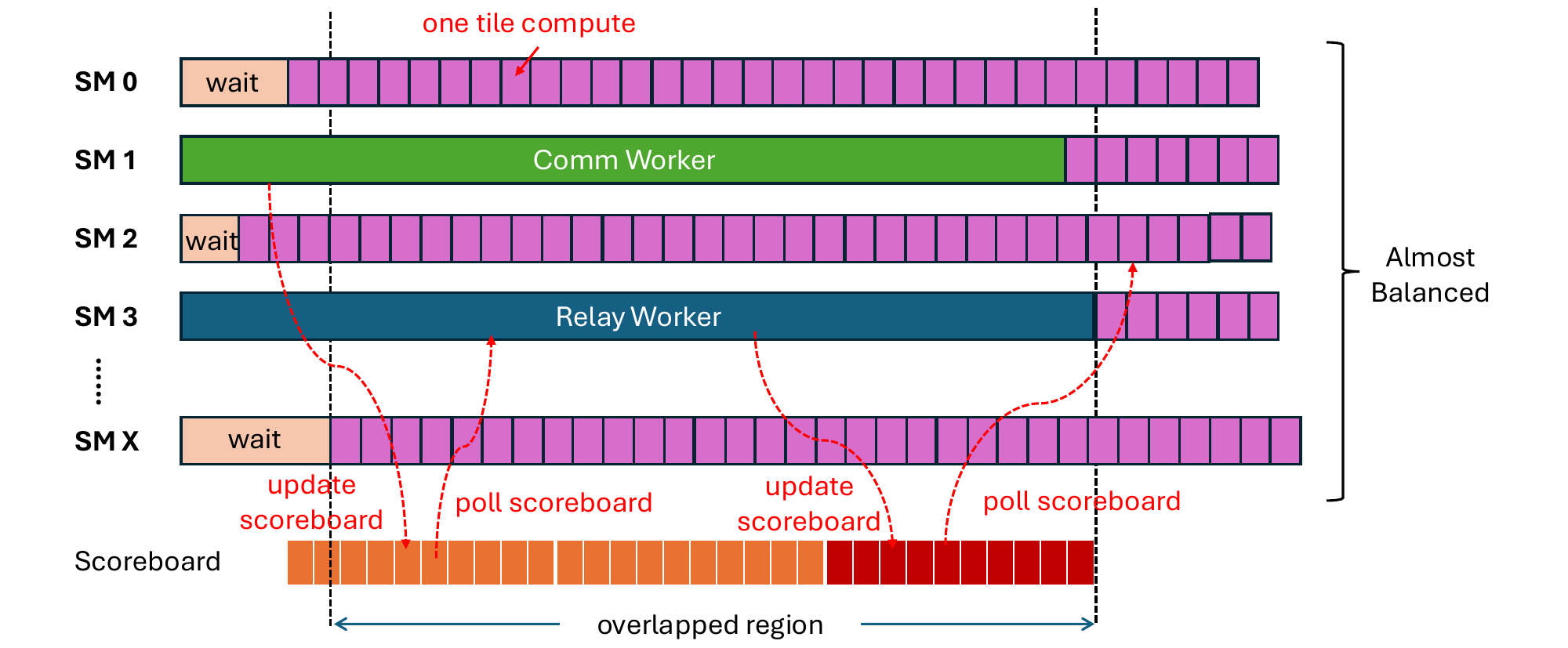}
    \caption{Example timeline and workflow of Dispatch+GroupGEMM MegaKernel.}
    \label{fig:megakernel}
\end{figure}

This mechanism naturally balances the workload: initially, all SMs process communication. As data arrives and the scoreboard fills, SMs transition to computation roles. Figure~\ref{fig:megakernel} visualizes this timeline: an \textit{overlapped region} naturally gives way to a mixed phase where Comm, Relay, and Comp workers operate in parallel, maximizing system throughput.

\textbf{Bandwidth Optimization via Relay Workers}
While the Relay Worker facilitates synchronization, its primary utility lies in bandwidth optimization via {intra-rank multicast}.
In standard MoE routing, a token may be routed to multiple experts residing on the same destination GPU (e.g., Top-8 routing where both experts are on Rank $i$). Naive approaches send the token multiple times over NVLink.

\ours{} optimizes this by transmitting the token only once to the destination rank. The Relay Worker on the destination rank then detects this condition and performs a local copy to replicate the token for different experts. This replaces expensive inter-rank bandwidth with high-speed HBM bandwidth.

\begin{table}[h]
\centering
\caption{Probability analysis of bandwidth reduction (Top-8, 8 GPUs). $X$ denotes distinct destination ranks.}
\label{tab:prob_saving}
\begin{tabular}{c c c c}
\toprule
\textbf{Distinct Ranks ($X$)} & \textbf{Saved Sends} & \textbf{Probability $P(X)$} & \textbf{Formula} \\
\midrule
1 & 7 & $4.8 \times 10^{-7}$ & $\binom{8}{1} 1! S_2(8,1)/8^8$ \\
2 & 6 & $4.2 \times 10^{-4}$ & $\binom{8}{2} 2! S_2(8,2)/8^8$ \\
3 & 5 & $1.9 \times 10^{-2}$ & $\binom{8}{3} 3! S_2(8,3)/8^8$ \\
4 & 4 & $0.170$ & $\binom{8}{4} 4! S_2(8,4)/8^8$ \\
5 & 3 & $0.420$ & $\binom{8}{5} 5! S_2(8,5)/8^8$ \\
6 & 2 & $0.320$ & $\binom{8}{6} 6! S_2(8,6)/8^8$ \\
7 & 1 & $0.067$ & $\binom{8}{7} 7! S_2(8,7)/8^8$ \\
8 & 0 & $0.002$ & $\binom{8}{8} 8! S_2(8,8)/8^8$ \\
\bottomrule
\end{tabular}%
\end{table}

We quantify the theoretical benefit of this optimization in Table~\ref{tab:prob_saving}. Assuming Top-$k=8$ and a balanced distribution across 8 GPUs, the expected number of distinct destination ranks is significantly lower than $k$.
The expectation calculation reveals that for Top-8 routing, we only need to transmit to an average of $5.25$ unique ranks, resulting in a theoretical traffic reduction of $\approx 34\%$.

\subsection{GroupGEMM+Combine MegaKernel Design}

We apply the same MegaKernel principles, including dynamic SM scheduling and scoreboard synchronization, to GroupGEMM+Combine.
The workflow involves Comp-Workers executing the GEMM and signalling Comm-Workers. The Comm-Workers transmit partial results to their source ranks. Finally, a {Reduce-Worker} (analogous to the Relay Worker) aggregates the incoming data.

A crucial distinction in this phase is the {Top-$k$ accumulation constraint}. To ensure numerical correctness during the reduction, the Reduce-Worker must wait until {all} Top-$k$ contributions for a specific token have arrived before performing the summation. Pre-mature reduction (accumulating on local ranks before sending) would violate the non-associativity of floating-point arithmetic, leading to non-bitwise {identical} results. By enforcing this strict barrier in the Reduce-Worker logic, \ours{} maintains bitwise consistency with the baseline while still hiding the latency of the underlying transfers.
\section{Optimization Space and AutoTuning}
\label{sec:autotuning}

\begin{table}[t]
\centering
\caption{Notations and Symbols for Perf Model}
\label{tab:notation}
\begin{tabular}{l l}
\toprule
\textbf{Symbol} & \textbf{Description} \\
\midrule
\multicolumn{2}{c}{\textit{Hardware Specifications ($H$)}} \\
\midrule
$N_{SM}$ & Total number of Streaming Multiprocessors \\
$\mathcal{P}_{peak}$ & Peak compute performance (TFLOPS) \\
$\beta_{HBM/NVL}$ & Peak bandwidth for HBM / NVLink (GB/s) \\
$w_{sat}$ & Warps to saturate bandwidth (e.g., 1024) \\
$\tau_{sync}$ & Synchronization overhead ($\approx 2\mu s$) \\
\midrule
\multicolumn{2}{c}{\textit{Problem \& Constant Configs ($P$)}} \\
\midrule
$B_M, B_N$ & GEMM block dimensions (e.g., 128) \\
$\mu$ & GEMM FLOPs utilization, related to warps $w$ \\
$N_{tok}$ & Tokens per rank \\
$H_{dim}$ & Hidden dimension size \\
$H_{inter}$ & Intermediate dimension size \\
$S_{tok}$ & Bytes per token \\
$V_{nvl}, V_{hbm}$ & Total data volume for NVLink/HBM \\
\midrule
\multicolumn{2}{c}{\textit{Optimization Variables ($C \in \mathcal{S}$)}} \\
\midrule
$w_{disp/comb}$ & Warps per Worker \\
$N_{disp/comb}$ & SMs for Comm-Workers \\
$N_{relay/red}$ & SMs for Relay/Reduce Workers \\
\midrule
\multicolumn{2}{c}{\textit{Latency Functions}} \\
\midrule
$t_{up}, t_{down}$ & Latency of single Up/Down GEMM block \\
$L_{disp/comb}$ & Latency of full Comm stage \\
$L_{swiglu}$ & Latency of SwiGLU kernel \\
\bottomrule
\end{tabular}
\end{table}

The \ours{} MegaKernel exposes a complex configuration space that governs the fine-grained interplay between computation, communication, and memory access. Relying on manual heuristics is insufficient due to the diversity of model sizes and hardware settings. In this section, we formalize these parameters into a unified search space and propose a rigorous analytical performance model to automate the selection of optimal configurations. Furthermore, we introduce a priority-based scheduling mechanism to maximize the efficacy of overlap.
Throughout this Section, we use symbols from Table~\ref{tab:notation} to explain our techniques.

\subsection{Tunable Primitives and Trade-offs}

We classify the optimization configs for the Dispatch+ GroupGEMM stage into three primitives: Warp Allocation ($w_{disp}$), Communication Worker Count ($N_{disp}$), and Relay Worker Count ($N_{relay}$). A symmetric set ($w_{comb}, N_{comb}, N_{red}$) applies to the GroupGEMM+ Combine stage. We analyze the critical system trade-offs for each.

\textbf{Warp Allocation per Worker ($w_{disp}, w_{comb}$).}
The number of warps allocated to each SM is a global parameter that creates a fundamental trade-off between compute stability and communication throughput. The valid search space is typically $\{8, 16, 32\}$.
From a compute perspective, for the Comp-Worker, empirical analysis on Hopper GPUs shows that 8 warps are sufficient to saturate the Tensor Core pipeline. Allocating excessive warps (e.g., 32) forces the warp scheduler to issue instructions aggressively, leading to high power density. This can trigger hardware thermal throttling, reducing the clock frequency and degrading overall FLOPS.
From a communication perspective, conversely, Comm-Workers are latency-bound. They benefit significantly from higher warp counts (32 warps), which provide enough instruction-level parallelism (ILP) to hide memory access latency and saturate the NVLink bandwidth.
Since the MegaKernel requires a uniform grid configuration, $w$ must be tuned to balance the risk of thermal throttling against the need for bandwidth saturation.

\textbf{Communication Worker Count ($N_{disp}, N_{comb}$).}
This parameter controls the parallelism of the communication stage and determines the resource partition between producers and consumers. The valid range is $[1, N_{SM})$.
There are two types of resutls when tuning this parameter. The fist is {over-allocation.} If $N_{disp}$ is too large (approaching $N_{SM}$), the kernel begins with a communication burst where almost all SMs act as producers. This blocks the consumer side, preventing the early start of Comp-Workers and effectively serializing the execution.
The second is {under-allocation.} If $N_{disp}$ is too small, the system becomes communication-bound. The compute workers drain the scoreboard faster than producers can fill it, leading to idle compute SMs.
The optimal $N_{disp}$ shifts dynamically with the arithmetic intensity of the workload and must be tuned per problem size.

\textbf{Relay Worker Count ($N_{relay}, N_{red}$).}
As introduced in Section~\ref{sec:design}, Relay Workers enable {intra-rank multicast} to reduce inter-rank traffic. However, this optimization comes with a resource cost.
Standard EP implementations utilize either \texttt{AllGather} or \texttt{AllToAll}. The communication volume for \texttt{AllGather} is inflated by the world size $W$: $ W \times N_{tok} \times S_{tok}$, where $N_{tok}$ is the token count per rank and $S_{tok}$ is bytes per token. Conversely, \texttt{AllToAll} transmits inflated data based on the Top-$k$ factor: $N_{tok} \times topk \times S_{tok}$.
\ours{} optimizes this to approximately $N_{tok} \times 5.25 \times S_{tok}$ (for Top-8). To realize this gain, Relay Workers must consume SM resources to process scoreboard signals and perform HBM copying.
On one hand, there is a strict constraint: 
$$N_{disp} + N_{relay} < N_{SM}$$
Violating this risks deadlock, as there may be insufficient SMs to schedule the Relay Workers required to unlock the computation.
On the other hand, we must allocate enough Relay Workers to ensure they do not become the bottleneck (blocking Comm-Workers), while minimizing their count to preserve SMs for GroupGEMM. The search range is typically $[1, N_{SM} - N_{disp})$.

\subsection{Analytical Performance Model}

Our optimization space is defined by the tuple 
$$\mathcal{C} = (w_{disp}, N_{disp}, N_{relay}, w_{comb}, N_{comb}, N_{red})$$
We construct an analytical model to predict the end-to-end latency $L_{total}$ for any given $\mathcal{C}$, hardware specification $H$, and problem size $P$. Table~\ref{tab:notation} summarizes the symbols used.

\begin{algorithm}[t]
\caption{Performance Model \& Config Search}
\label{algo:perf_model}
\begin{algorithmic}[1]
\Require Problem $P$, Hardware $H$, Config Space $\mathcal{S}$
\Ensure Optimal Config $C^*$, Min Latency $L_{min}$

\State $L_{min} \gets \infty$; $C^* \gets \emptyset$
\ForAll{$C \in \mathcal{S}$}
    \State \Comment{\textbf{1. Basic Operation Latencies}}
    \State $t_{up} \gets \text{CalcGEMM}(H, P, \text{up})$
    \State $t_{down} \gets \text{CalcGEMM}(H, P, \text{down})$
    \State $L_{swiglu} \gets \text{CalcSwiglu}(P, H)$
    
    \State \Comment{\textbf{2. Stage 1: Dispatch + UpGEMM Overlap}}
    \State $N_{comp1} \gets H.N_{SM} - C.N_{disp}$
    \State $L_{disp} \gets \text{CalcDispLat}(P, H, C.N_{disp})$
    \State $L_{up} \gets \text{CalcCompLat}(P, t_{up}, N_{comp1})$
    
    \If{$L_{up} > L_{disp}$}
        \State $L_{S1} \gets L_{disp} + (L_{up} - L_{disp}) \times \frac{H.N_{SM}}{N_{comp1}}$
    \Else
        \State $L_{S1} \gets L_{disp} + t_{up}$
    \EndIf

    \State \Comment{\textbf{3. Stage 2: DownGEMM + Combine Overlap}}
    \State $N_{comp2} \gets H.N_{SM} - C.N_{comb}$
    \State $L_{comb}, t_{red} \gets \text{CalcCombLat}(P, H, C.N_{comb})$
    \State $L_{down} \gets \text{CalcCompLat}(P, t_{down}, N_{comp2})$
    
    \State \Comment{Estimate Reduce overlap (filling the gap)}
    \State $L_{base} \gets \max(L_{down}, L_{comb})$
    \State $w_{gap} \gets |L_{down} - L_{comb}| \times N_{comp2}$ 
    \State $w_{red} \gets C.N_{red} \times t_{red}$ 
    \State $w_{rem} \gets \max(0, w_{red} - w_{gap})$
    \State $L_{S2} \gets L_{base} + w_{rem} / H.N_{SM}$

    \State \Comment{\textbf{4. Update Optimal}}
    \State $L_{total} \gets L_{S1} + L_{S2} + L_{swiglu}$
    \If{$L_{total} < L_{min}$}
        \State $L_{min} \gets L_{total}$
        \State $C^* \gets C$
    \EndIf
\EndFor
\State \Return $C^*, L_{min}$
\end{algorithmic}
\end{algorithm}

\paragraph{Effective Bandwidth.} 
We define the effective bandwidth function $\mathcal{B}(n, \beta)$ given $n$ active SMs and peak bandwidth $\beta$. Let $w$ be the number of warps per SM and $w_{sat}$ (e.g., 1024) be the saturation threshold. The available bandwidth is limited by either the hardware peak or the number of active warps:
\begin{equation}
    \mathcal{B}(n, \beta) = \min\left( n \cdot w \cdot \frac{\beta}{w_{sat}}, \beta \right)
\end{equation}

\paragraph{GEMM Block Latency (\texttt{CalcGEMM})}
This function estimates the execution time of a single GroupGEMM tile. For a tile size $B_M \times B_N$ and reduction dimension $K$, the latency is derived from peak FLOPS scaled by MFU $\mu$ {(the ratio of observed FLOPS to theoretical peak)}. As each tile needs to wait for a signal or set a signal, we add a synchronization overhead $\tau_{sync}$.
\begin{equation}
    t_{gemm}(K) = \frac{2 \cdot B_M \cdot B_N \cdot K}{\mathcal{P}_{peak} \cdot (\mu / N_{SM})} + \tau_{sync}
\end{equation}
Consequently, $t_{up} = t_{gemm}(H_{dim})$ and $t_{down} = t_{gemm}(H_{inter})$, where $H_{dim}$ is hidden size and $H_{inter}$ is MoE intermediate size.
Note that $\mu$ is related to warps $w$. We set the following values according to profiling:
$$\mu(w=8) = 0.7 \ \ \mu(w=16) = 0.65 \ \ \mu(w=32)=0.6$$

\paragraph{SwiGLU Latency (\texttt{CalcSwiglu}).}
SwiGLU is strictly memory-bound. For $N_{tok}$ tokens of size $S_{tok}$:
\begin{equation}
    L_{swiglu} = \frac{2 \cdot N_{tok} \cdot S_{tok}}{\beta_{HBM}}
\end{equation}

\paragraph{Dispatch Latency (\texttt{CalcDispLat}).}
The dispatch latency sums the costs of NVLink transfers (remote) and HBM writes (local/relay). With $N_{disp}$ and $N_{relay}$ active communication and relay SMs:
\begin{equation}
    L_{disp} = \frac{V_{nvl}}{\mathcal{B}(N_{disp}, \beta_{NVL})} + \frac{V_{hbm}}{\mathcal{B}(N_{relay}, \beta_{HBM})}
\end{equation}

\paragraph{Computation Latency (\texttt{CalcCompLat}).}
The total computation time for a stage is determined by the number of waves required to process all tiles on the allocated compute SMs ($N_{comp}$). Let $N_{tiles}$ be the total number of GEMM blocks:
\begin{equation}
    L_{comp}(t_{block}, N_{comp}) = \left\lceil \frac{N_{tiles}}{N_{comp}} \right\rceil \cdot t_{block}
\end{equation}

\paragraph{Combine Latency (\texttt{CalcCombLat}).}
Similar to Dispatch, this function calculates the scatter latency via NVLink. It also returns $t_{red}$, the time for a single SM to perform the reduction of incoming tokens in HBM:
\begin{align}
    L_{comb} &= \frac{V_{nvl}}{\mathcal{B}(N_{comb}, \beta_{NVL})} \\
    t_{red} &= \frac{V_{hbm\_total}}{\mathcal{B}(1, \beta_{HBM})}
\end{align}

\paragraph{Overlap Simulation Logic.}
Algorithm~\ref{algo:perf_model} integrates these equations. Lines 10-14 capture the primary overlap: if computation ($L_{up}$) dominates communication ($L_{disp}$), the effective latency ($L_{S1}$) is the communication time plus the tail of the computation. Lines 22-26 model the gap filling for the Combine phase: if the reduction work ($w_{red}$) exceeds the bubble between GroupGEMM and Combine ($w_{gap}$), the excess work ($w_{rem}$) extends the critical path.

\subsection{Priority-based Token Scheduling}
In previous performance model, we have a strong assumption that computation and communication can perfectly overlap. To achieve this, we need additional optimizations.
While the mapping algorithm (Algorithm~\ref{alg:token_mapping}) determines {where} tokens go, the {order} of transmission determines the efficiency of overlap. In a naive implementation, a Comm-Worker might transmit tokens in a natural order (e.g., for Expert 0, then Expert 1, ...). However, the remote Computation Worker is deterministic: it processes tiles for Expert 0, then Expert 1, etc.
If the data for local first expert arrives late, the Comp-Worker stalls, even if data for later experts is already available in HBM. This head-of-line blocking negates the benefits of overlap.

To resolve this, we enforce strict alignment between production and consumption. The Comp-Workers process experts in ascending order (locally $E_0, E_1, \dots$). Correspondingly, Communication Workers sort their transmission queues to prioritize tokens destined for remote Expert 0, followed by Expert 1. Since the global offsets are pre-calculated via prefix sum, this sorting is achieved implicitly by iterating through the expert offset array, ensuring zero-overhead prioritization.
\section{Implementation}
\label{sec:implementation}

Implementing the \ours{} MegaKernel requires managing massive concurrency, complex dependency chains, and low-level hardware intrinsics. Attempting such a fusion in raw CUDA would entail prohibitive engineering complexity, likely exceeding 100k lines of code and becoming unmaintainable.
To mitigate this, we leverage and extend {Triton-Distributed}~\cite{triton-dist}, a language framework that exposes block-level and warp-level control while abstracting standard boilerplate. The core logic of \ours{} is implemented in approximately 21k lines of Python code (an estimated 5-10$\times$ reduction compared to a theoretical CUDA equivalent).

\subsection{Framework Extensions}
Triton-Distributed serves as the backbone for our implementation, offering three critical primitives that enable the MegaKernel design:

\noindent
\textbf{Native NVSHMEM Support:} It exposes the entire NVSHMEM API surface, allowing us to write distributed kernels where pointers reference remote GPU memory directly.

\noindent
\textbf{Memory-Semantic Signals:} It provides \texttt{ld\_acquire} and \texttt{st\_release} intrinsics. These are essential for our scoreboard mechanism, ensuring that signal updates in global memory are visible across SMs with correct ordering constraints.

\noindent
\textbf{Warp-Level Control:} We utilize extensions to Triton's SPMD model that allow accessing `warp\_id` and performing intra-warp synchronization. This enables us to implement the {Comm-Worker} logic where threads within a warp collaborate together.

\subsection{MegaKernel Structure}
The architectural skeleton of the MegaKernel follows a dynamic fetching loop. As shown in Listing~\ref{lst:megakernel}, each persistent thread block (SM) continuously fetches tasks from a global atomic counter until the queue is drained. This implementation allows seamless transitions between roles (e.g., from Dispatch to Computation) without kernel relaunch overhead.

\definecolor{codegreen}{rgb}{0,0.6,0}
\definecolor{codegray}{rgb}{0.5,0.5,0.5}
\definecolor{codepurple}{rgb}{0.58,0,0.82}
\definecolor{backcolour}{rgb}{0.95,0.95,0.92}

\lstset{
    backgroundcolor=\color{backcolour},   
    commentstyle=\color{codegreen},
    keywordstyle=\color{magenta},
    numberstyle=\tiny\color{codegray},
    stringstyle=\color{codepurple},
    basicstyle=\ttfamily\footnotesize,
    breakatwhitespace=false,         
    breaklines=true,                 
    captionpos=b,                    
    keepspaces=true,                 
    numbers=left,                    
    numbersep=5pt,                  
    showspaces=false,                
    showstringspaces=false,
    showtabs=false,                  
    tabsize=2,
    language=Python
}

\begin{lstlisting}[caption={Pseudocode of the UniEP MegaKernel Loop}, label={lst:megakernel}]
@triton.jit
def dispatch_group_gemm_kernel(
    # ... ptrs and stride args ...
    LOCK_PTR, TASK_COUNT
):
    sm_id = tl.program_id(0)
    # Dynamic Task Fetching
    task_id = tl.atomic_add(LOCK_PTR, 1)
    
    while task_id < TASK_COUNT:
        task_type, task_args = decode_task(task_id)
        
        if task_type == TASK_GROUPGEMM:
            # Perform Computation
            compute_tile(task_args, ...)
            
        elif task_type == TASK_DISPATCH:
            # Perform Communication via NVSHMEM
            dispatch_tokens(task_args, ...)
            
        elif task_type == TASK_RELAY:
            # Handle Signals and Local Copy
            relay_signal(task_args, ...)
            
        # Fetch next task
        task_id = tl.atomic_add(LOCK_PTR, 1)
\end{lstlisting}

\subsection{Worker Micro-Architecture}

\textbf{Communication Worker.}
We implement inter-GPU data movement using NVSHMEM intrinsics. We specifically prioritize the \texttt{putmem\_warp} primitive over \texttt{getmem}. Empirical testing reveals that push-based semantics (\texttt{put}) yield consistently higher bandwidth utilization on NVLink interconnects compared to pull-based (\texttt{get}) approaches. Each warp is responsible for moving one tokens. Synchronization with {Relay Workers} is enforced via \texttt{st\_release}, guaranteeing that data payload writes complete before the signal becomes visible.

\textbf{Computation Worker.}
The GroupGEMM logic is implemented using high-level Triton \texttt{tl.dot} and \texttt{tl.load} operations. Although we do not currently utilize the Tensor Memory Accelerator (TMA) hardware features of Hopper, our implementation remains highly efficient. For a standard tile size of $128 \times 256 \times 64$, the compute loop execution time is approximately $22\mu s$, achieving 60\%--70\% of the theoretical Tensor Core peak performance. This is comparable to heavily optimized CUTLASS~\cite{cutlass} kernels.

\textbf{Relay \& Reduce Workers.}
These workers operate with high warp-level parallelism. In the {Reduce} phase, each warp manages the aggregation of a specific token across its Top-$k$ replicas. By utilizing vectorized load/store and bypassing shared memory, we minimize the latency overhead of the reduction step.

\subsection{Runtime Integration and Optimization}

We integrated \ours{} into the forward and backward passes of standard MoE training loops. However, the runtime overhead of the performance model presented a challenge.
The search space $\mathcal{S}$ contains approximately $10^5$ configurations. {We iteratively enumerate all the configrations in parallel and predict the latency. Finally, the configuration that produces the lowest latency is kept.} Our initial Python implementation required $\approx 100$ seconds to traverse this space, which is unacceptable for online tuning.

To resolve this, we reimplemented the search algorithm in C++ parallelized with OpenMP. This optimized solver completes the search in roughly $144$ ms. Furthermore, we employ a bucketing memoization strategy. We discretize the input sequence length $N_{tok}$ into 4096-token buckets. The performance model is queried only when the workload crosses a bucket boundary. The optimal configuration is then cached and reused for subsequent iterations. This reduces the amortized overhead of the auto-tuner to negligible levels in long-running training jobs.

\section{Evaluation}
\label{sec:evaluation}

In this Section, we evaluate the performance of \ours{} across diverse hardware configurations and realistic model architectures. Our evaluation aims to answer three key questions:
\begin{enumerate}
    \item Can the proposed performance model accurately identify optimal configurations in a vast search space?
    \item How does our MegaKernel compare against state-of-the-art baselines in kernel-level benchmarks?
    \item What is the forward and backward performance gain for MoE?
\end{enumerate}

\subsection{Evaluation Setup}

\textbf{Hardware Environment.} We conduct experiments on two NVIDIA Hopper GPU clusters with distinct interconnect characteristics, as detailed in Table~\ref{tab:hardware_config}. {Cluster 1} represents a bandwidth-constrained environment (200 GB/s/dir NVLink), typical of older generation interconnects, while {Cluster 2} represents a high-bandwidth environment (400 GB/s/dir NVLink). Both clusters are equipped with 8 GPUs per node.

\begin{table}[h]
\centering
\caption{Hardware Specifications of Experimental Clusters}
\label{tab:hardware_config}
\begin{tabular}{l c c c c c}
\toprule
\textbf{Cluster} & \textbf{GPUs} & \textbf{HBM Cap.} & \textbf{HBM BW} & \textbf{NVL BW (Uni)} & \textbf{Peak BF16} \\
\midrule
Cluster 1 & 8 & 80 GB & 3.35 TB/s & 200 GB/s & 989 TFLOPS \\
Cluster 2 & 8 & 96 GB & 3.90 TB/s & 400 GB/s & 148 TFLOPS \\
\bottomrule
\end{tabular}%
\end{table}

\textbf{Workloads.} We select 12 representative MoE configurations from production models, including the DeepSeek~\cite{deepseek-v3, deepseek-v2, deepseek-ocr2, deepseek-r1}, Qwen \cite{qwen3}, and Kimi~\cite{kimi-k2, kimi-linear} families. These configurations cover a wide range of expert counts (64-512) and hidden dimensions, as listed in Table~\ref{tab:model_shapes}. We evaluate sequence lengths of 8k, 32k, and 128k to assess performance across varying training settings.

\begin{table}[t]
\centering
\caption{MoE Model Configurations Evaluated}
\label{tab:model_shapes}
\begin{tabular}{l l c c c c}
\toprule
\textbf{ID} & \textbf{Model Name} & \textbf{$H_{dim}$} & \textbf{$H_{inter}$} & \textbf{$N_{exp}$} & \textbf{Top-$k$} \\
\midrule
MoE-1 & DeepSeek-MoE-16B & 2048 & 1408 & 64 & 6 \\
MoE-2 & DeepSeek-OCR-2 & 1280 & 896 & 64 & 6 \\
MoE-3 & DeepSeek-V2-Lite & 2048 & 1408 & 64 & 6 \\
MoE-4 & DeepSeek-V2-Chat & 5120 & 1536 & 160 & 6 \\
MoE-5 & DeepSeek-R1 & 7168 & 2048 & 256 & 8 \\
MoE-6 & Qwen3-30B-A3B & 2048 & 768 & 128 & 8 \\
MoE-7 & Qwen3-235B-A22B & 4096 & 1536 & 128 & 8 \\
MoE-8 & Qwen3-Coder-480B & 6144 & 2560 & 160 & 8 \\
MoE-9 & Qwen3-Next-80B & 2048 & 512 & 512 & 10 \\
MoE-10 & Qwen3-Omni-30B & 1024 & 384 & 128 & 6 \\
MoE-11 & Kimi-K2 & 7168 & 2048 & 384 & 8 \\
MoE-12 & Kimi-Linear-48B & 2304 & 1024 & 256 & 8 \\
\bottomrule
\end{tabular}%
\end{table}

\begin{table}[h]
\centering
\caption{Optimal Optimization Configurations Identified by Performance Model on Cluster 2 (SeqLen 32k)}
\label{tab:auto_tuning}
\begin{tabular}{l c c c c c c}
\toprule
\textbf{Model ID} & \textbf{$N_{disp}$} & \textbf{$N_{comb}$} & \textbf{$N_{relay}$} & \textbf{$N_{red}$} & \textbf{$warps$} & {\textbf{Tune Time (ms)}} \\
\midrule
MoE-1  & 12 & 20 & 5 & 78 & 32 & 29.6 \\
MoE-2  & 16 & 28 & 1 & 78 & 32 & 29.7 \\
MoE-3  & 12 & 20 & 5 & 78 & 32 & 30.0 \\
MoE-4  & 12 & 20 & 5 & 78 & 32 & 46.4\\
MoE-5  & 8  & 16 & 1 & 78 & 32 & 80.5 \\
MoE-6  & 16 & 28 & 5 & 78 & 32 & 49.6\\
MoE-7  & 12 & 20 & 5 & 78 & 32 & 49.5 \\
MoE-8  & 8  & 12 & 1 & 78 & 32 & 58.2 \\
MoE-9  & 20 & 32 & 5 & 78 & 32 & 149.9 \\
MoE-10 & 28 & 40 & 5 & 78 & 32 & 41.5 \\
MoE-11 & 8  & 16 & 1 & 78 & 32 & 144.3 \\
MoE-12 & 16 & 24 & 1 & 78 & 32 & 80.1 \\
\bottomrule
\end{tabular}%
\end{table}


\textbf{Baselines.} We compare \ours{} against two baselines:
\textbf{Serial} (DeepEP~\cite{deepep} + TransformerEngine~\cite{te}): A non-overlapping baseline using state-of-the-art kernels. Communication is handled by DeepEP (optimized  with NVSHMEM), and computation by TransformerEngine. We install the latest DeepEP (commit hash 3fcf25) and TransformerEngine v2.11.
\textbf{COMET:} The current state-of-the-art overlapping solution used in enterprise training. It employs dual streams to pipeline NVLink communication with GroupGEMM computation. Its communication is done via copy engine (DMA engines), which is AllGather, while the computation part is implemented by CUTLASS~\cite{cutlass}. We use commit hash 19831c.

\subsection{Optimization Configs and Model Accuracy}

{\textbf{Optimization Space Size.}
The optimization configuration space $\mathcal{C} = (N_{disp}, N_{relay}, N_{comb}, N_{red}, w)$ is defined as follows. Both $N_{disp}$ and $N_{comb}$ are constrained to multiples of 4, i.e., $N_{disp}, N_{comb} \in \{4, 8, 12, \ldots, N_{SM}\}$, yielding $\lfloor \frac{N_{SM}}{4} \rfloor = 33$ choices each on Cluster~1 (132 SMs). $N_{relay}$ ranges from 1 to $\frac{N_{disp}}{2}$ in steps of 4, giving on average 8 choices. $N_{red}$ ranges from 1 to $N_{SM}$ in steps of 16, giving 8 choices. The warp count $w \in \{8, 16, 32\}$ provides 3 choices. The total search space size is therefore of size $33\times 33 \times 8 \times 8 \times 3 = 209,088$ (roughly $10^5$).}

\textbf{Tuning Results and Overhead.}
Table~\ref{tab:auto_tuning} showcases selected optimal configurations for Cluster 2 (SeqLen 32k).
The columns $N_{disp}$, $N_{comb}$, $N_{relay}$, $N_{red}$, $warps$ represent the number of SMs for Dispatch Comm-Workers, Combine Comm-Workers, Relay Workers, Reduce Workers and warps per worker. We find that the optimal configurations uniformly use all the SMs for final reduction in GroupGEMM+Combine. And the warps for both dispatch and combine are 32.

{We also show the tuning time for each configuration in Table~\ref{tab:auto_tuning}. The tuning time ranges from 29.7ms to 149.9ms. The tuning time varies with MoE configurations because the number of tiles is affected by input shapes. More tiles require more time to calculate predicted latency using our performance model. 
The auto-tuning procedure is a one-time cost: once the optimal configuration is identified for a given hardware and model shape, it is cached and reused across all subsequent training iterations.
}

\textbf{Model Accuracy.}
To validate our performance model (Section~\ref{sec:autotuning}), we compare the predicted optimal configurations against ground-truth results. We performed an exhaustive search (taking hours) for configurations including Kimi-K2, Qwen3-235B, and Qwen3-30B. Comparing these against our performance model's output (taking $\approx 144$ms, {the longest for Kimi-K2}), the real absolute latency error ranges between \textbf{0.5\% and 6.5\%}, average $3.8\%$. {We note that our analytical model tends to yield optimistic latency estimates, as cache contention at runtime can degrade performance in ways that are not yet well captured by the model.} This high fidelity confirms that our analytical model effectively captures the complex hardware behaviors. 

\subsection{Numeric Precision Analysis}

{\ours{} guarantees the determinism of token send and receive ordering through the deterministic token mapping described in Section~\ref{sec:dispatch_groupgemm}. This ensures that the accumulation order of the Transposed GroupGEMM used for weight gradient computation in the backward pass is likewise preserved. We conduct two experiments analyzing the impact of deterministic token mapping on bitwise numerical equivalence and on performance.}

{\textbf{Bitwise Equivalence Verification.}
We compare COMET and \ours{} against a reference serial implementation based on PyTorch+NCCL with no overlap optimization. COMET does not guarantee that its internal token ordering is consistent with the no-overlap reference. We run all 12 MoE configurations from Table~\ref{tab:model_shapes} on Cluster~1 and record the element-wise maximum absolute difference (\texttt{max\_diff}) and the fraction of non-bitwise-equal output elements (\texttt{\%non-bw}). Results are presented in Table~\ref{tab:max-diff}.
COMET exhibits non-trivial numerical deviations from the reference: the maximum absolute difference reaches up to 0.25, with between 22\% and 29\% of output elements failing the bitwise equality check across all tested configurations. In contrast, \ours{} produces bitwise-identical results to the reference for every configuration in both forward and backward passes, directly validating the correctness of our deterministic token mapping scheme.}

\begin{table}[t]
\centering
\caption{{Numerical precision comparison against the no-overlap reference. `-' indicates the configuration is not supported by COMET.}}
\label{tab:max-diff}
\setlength{\tabcolsep}{5pt}
\begin{tabular}{lrrrr}
\toprule
& \multicolumn{2}{c}{\textbf{\ours{} (Ours)}} & \multicolumn{2}{c}{\textbf{COMET}} \\
\cmidrule(lr){2-3} \cmidrule(lr){4-5}
\textbf{ID} & \texttt{max\_diff} & \texttt{\%non-bw} & \texttt{max\_diff} & \texttt{\%non-bw} \\
\midrule
\midrule
MoE-1  & 0 & 0\% & 3.13e-02 & 21.69\% \\
MoE-2  & 0 & 0\% & -      & -     \\
MoE-3  & 0 & 0\% & 3.13e-02 & 21.69\% \\
MoE-4  & 0 & 0\% & 1.25e-01 & 22.74\% \\
MoE-5  & 0 & 0\% & 2.50e-01 & 29.17\% \\
MoE-6  & 0 & 0\% & 3.13e-02 & 29.14\% \\
MoE-7  & 0 & 0\% & 1.25e-01 & 28.98\% \\
MoE-8  & 0 & 0\% & 2.50e-01 & 29.19\% \\
MoE-9  & 0 & 0\% & 3.13e-02 & 29.14\% \\
MoE-10 & 0 & 0\% & 6.25e-02 & 29.05\% \\
MoE-11 & 0 & 0\% & 1.25e-01 & 29.31\% \\
MoE-12 & 0 & 0\% & - & - \\
\bottomrule
\end{tabular}
\end{table}

{\textbf{Performance Cost of Bitwise Reproducibility.}
Maintaining bitwise equivalence incurs a measurable performance cost, which manifests primarily in the backward pass. In the backward pass, this ordering constrains the reduction sequence of the Transposed GroupGEMM. Relaxing this constraint permits more aggressive overlap strategies. We can partition tokens into two sub-batches to better pipeline computation and communication at the cost of bitwise reproducibility.
Table~\ref{tab:bitwise-vs-non-bitwise} compares the latency of the bitwise-reproducible (\textbf{BW}) and non-bitwise (\textbf{NB}) variants of \ours{} on Cluster~1 at 32k sequence length. The forward pass is identical for both variants but the backward performance results are different. For most configurations, the non-bitwise variant delivers a 2\%--8\% speedup in end-to-end latency except two cases:
\textit{MoE-10  and MoE-11}: MoE-10 GroupGEMM arithmetic intensity is very low. Splitting into two sub-batches exposes a residual compute tail that was previously absorbed by the overlap, resulting in a slight performance regression.
MoE-11 GroupGEMM of is compute-heavy. Splitting into two sub-batches halves the number of SMs available per sub-batch, causing the GroupGEMM itself to slow down.}

\begin{table}[t]
\centering
\caption{{Performance comparison between \ours{} bitwise (\textbf{BW}) and non-bitwise (\textbf{NB}) variants on Cluster~1 (32k sequence length). Speedup ${>}1{\times}$ indicates NB is faster.}}
\label{tab:bitwise-vs-non-bitwise}
\setlength{\tabcolsep}{4pt}
\begin{tabular}{lrrrr}
\toprule
\textbf{ID} & \textbf{Fwd (ms)} & \textbf{BW Bwd (ms)} & \textbf{NB Bwd (ms)} & \textbf{Speedup} \\
\midrule
MoE-1  & 5.25  &  6.52 &  5.88 & 1.06$\times$ \\
MoE-2  & 4.46  &  4.19 &  3.67 & 1.06$\times$ \\
MoE-3  & 5.25  &  6.52 &  5.88 & 1.06$\times$ \\
MoE-4  & 9.97  & 15.14 & 13.92 & 1.05$\times$ \\
MoE-5  & 20.12 & 33.62 & 32.24 & 1.03$\times$ \\
MoE-6  & 5.00  &  6.06 &  5.88 & 1.02$\times$ \\
MoE-7  & 10.42 & 15.33 & 13.41 & 1.08$\times$ \\
MoE-8  & 20.16 & 31.83 & 30.79 & 1.02$\times$ \\
MoE-9  & 5.54  &  7.58 &  6.88 & 1.06$\times$ \\
MoE-10 & 4.59  &  3.55 &  3.83 & 0.97$\times$ \\
MoE-11 & 20.69 & 36.10 & 37.12 & 0.98$\times$ \\
MoE-12 & 6.23  &  8.18 &  7.13 & 1.08$\times$ \\
\bottomrule
\end{tabular}
\end{table}

\subsection{Kernel-level Performance}

\begin{figure*}[!t]
    \centering
    \includegraphics[width=\textwidth]{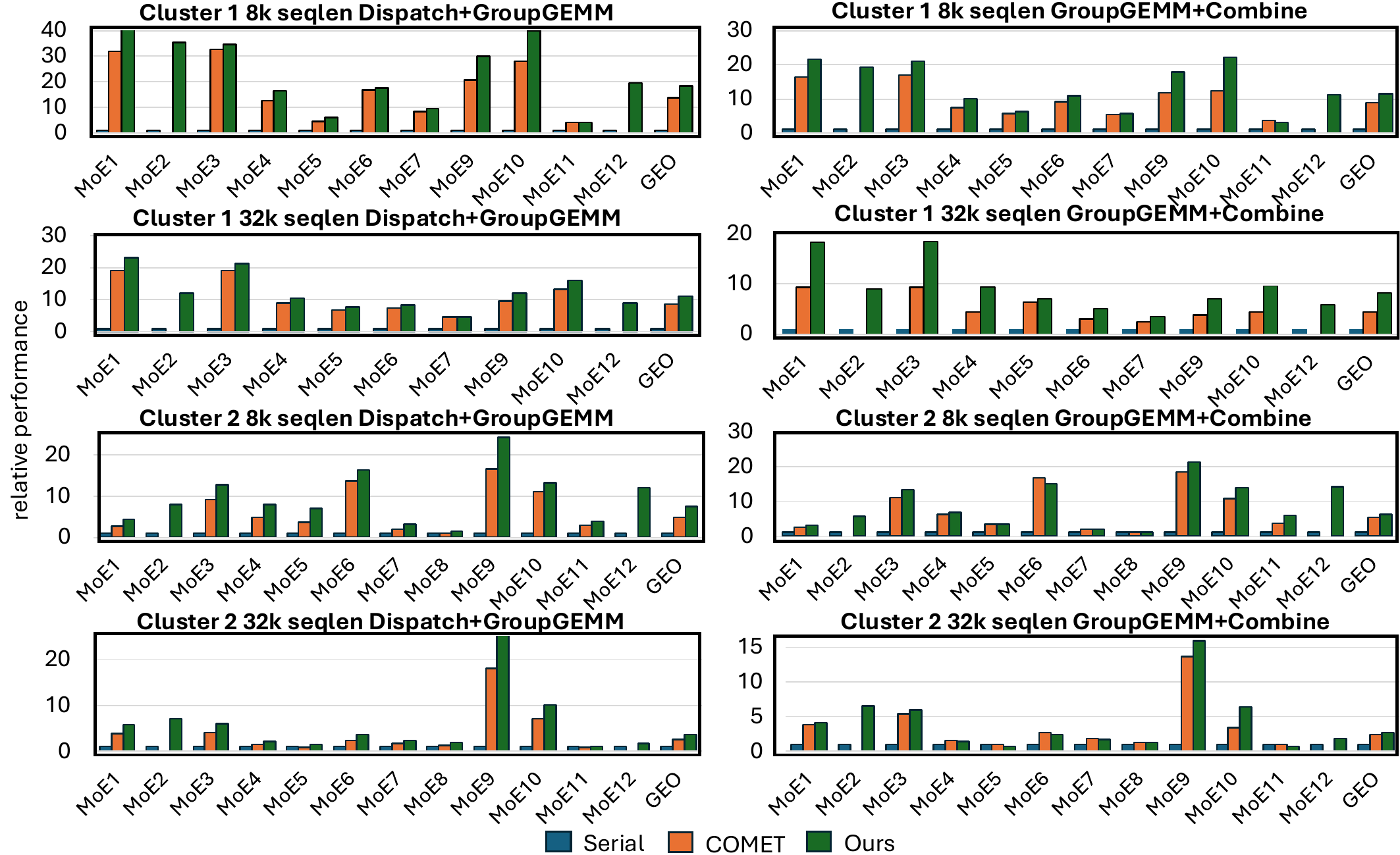}
    \caption{\ours{} kernel-level performance on Cluster 1 and Cluster 2.}
    \label{fig:kernel-level-perf}
\end{figure*}

\begin{figure*}[!t]
    \centering
    \includegraphics[width=\textwidth]{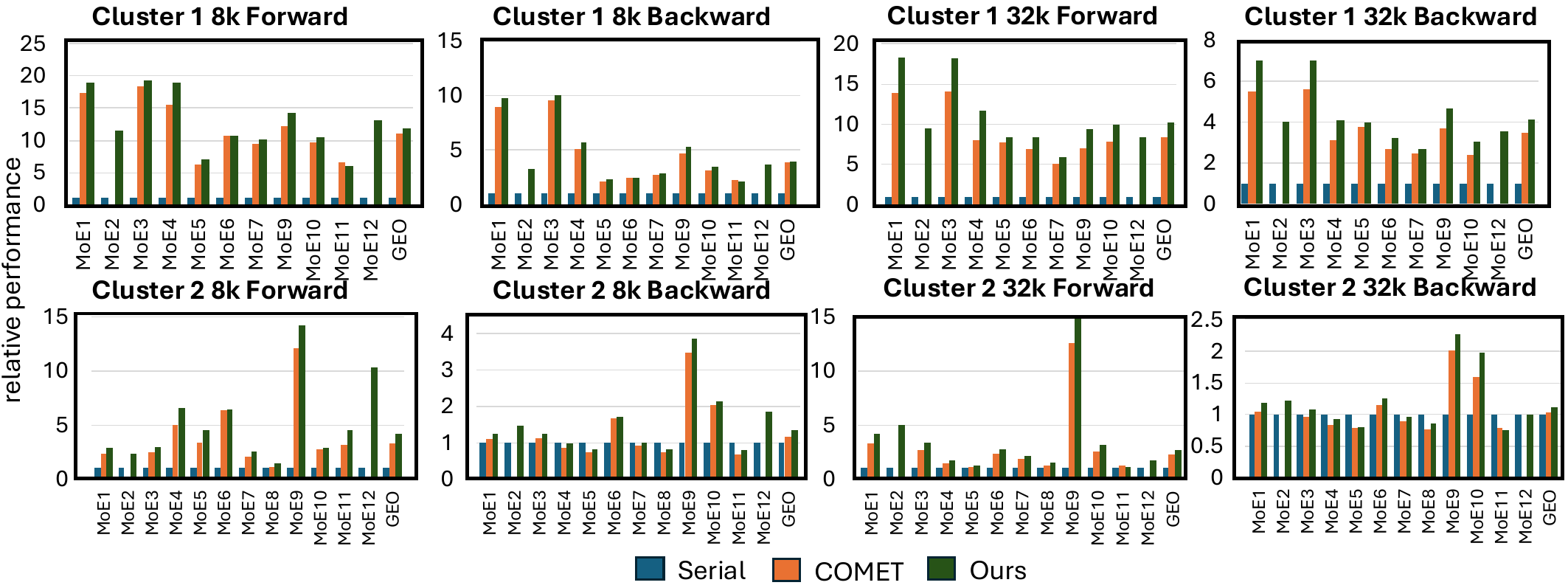}
    \caption{\ours{} layer-level performance on Cluster 1 and Cluster 2.}
    \label{fig:layer-level-perf}
\end{figure*}

We first evaluate the kernel-level performance of the Dispatch+ GroupGEMM and GroupGEMM+ Combine phases. Figure~\ref{fig:kernel-level-perf} summarizes the results.

\textbf{Dispatch+GroupGEMM.}
On {Cluster 1}, \ours{} achieves substantial gains. At 8k seqlen, we observe a geometric mean speedup of {18.40$\times$} over Serial and {1.32$\times$} over COMET. At 32k seqlen (excluding OOM cases MoE-8/11), the speedups are {11.20$\times$} (vs. Serial) and {1.30$\times$} (vs. COMET).
On {Cluster 2}, gains are slightly lower due to higher bandwidth availability but remain significant. At 8k seqlen, \ours{} outperforms Serial by 7.44$\times$ and COMET by 1.57$\times$. At 32k seqlen, speedups are 3.57$\times$ and 1.45$\times$, respectively.
COMET failed to run for MoE-2 and MoE-12 because COMET only implements kernels that specialize $H_{dim}$ and Top-$k$, $H_{dim}=$2304 and Top-$k$=6 are not supported.

\textbf{GroupGEMM+Combine.}
The trends are consistent. On Cluster 1, \ours{} achieves 11.56$\times$ (8k) and 8.23$\times$ (32k) speedups over Serial, and 1.31$\times$/1.87$\times$ over COMET. On Cluster 2, speedups over COMET are 1.21$\times$ (8k) and 1.07$\times$ (32k).

\textbf{Analysis of Baselines.}
The poor performance of the {Serial} baseline, despite using DeepEP, is attributed to the loop-over-experts pattern in standard GroupGEMM implementations (by TE). This requires frequent CPU-GPU synchronization to fetch shape metadata, causing massive overhead that dwarfs the optimized communication time.
Compared to {COMET}, \ours{} excels for two reasons: (1) Our Relay Workers reduce physical traffic via intra-node multicasting, which is not implemented in COMET. (2) Profiling traces reveal that COMET suffers from alignment bubbles, which are idle gaps caused by the asynchronous launch of kernels on dual streams. \ours{}, by fusing operations into a MegaKernel on one stream, eliminates these bubbles entirely.

\textbf{Cluster Sensitivity.}
Speedups are more pronounced on the bandwidth-constrained Cluster 1. This is expected: when the system is communication-bound, the efficacy of overlap is maximized. On Cluster 2, the primary gains shift from bandwidth hiding to synchronization overhead reduction.

\subsection{Layer-level Performance}

We aggregate the forward and backward pass latencies (excluding gating) in Figure~\ref{fig:layer-level-perf}.
On {Cluster 1} (8k SeqLen), \ours{} accelerates the forward pass by {11.98$\times$} vs. Serial and {1.08$\times$} vs. COMET. The backward pass shows 4.00$\times$ and 1.03$\times$ speedups.
On Cluster 1, our advantage over COMET grows with sequence length (1.22$\times$ forward speedup at 32k). This validates our {communication reduction optimization}: as token counts increase, the probability of redundant routing to the same rank increases, allowing our Relay Workers to save more NVLink bandwidth.
On {Cluster 2}, this traffic reduction effect is less dominant due to abundant bandwidth. Speedups over COMET stabilize at around 1.19$\times$ (forward, 32k) and 1.09$\times$ (backward, 32k).

\subsection{Long-Context Performance (128k)}

\begin{table}[t]
\centering
\caption{Absolute performance (ms) for 128k seqlen. `---' indicates unsupported config.}
\label{tab:longct}
\setlength{\tabcolsep}{2.5pt}
\begin{tabular}{l rrrrrr rrrrrr}
\toprule
& \multicolumn{6}{c}{\textbf{Cluster 1}} & \multicolumn{6}{c}{\textbf{Cluster 2}} \\
\cmidrule(lr){2-7} \cmidrule(lr){8-13}
\multirow{2}{*}{\textbf{ID}} & \multicolumn{2}{c}{\textbf{Serial}} & \multicolumn{2}{c}{\textbf{COMET}} & \multicolumn{2}{c}{\textbf{\ours}} & \multicolumn{2}{c}{\textbf{Serial}} & \multicolumn{2}{c}{\textbf{COMET}} & \multicolumn{2}{c}{\textbf{\ours}} \\
\cmidrule(lr){2-3} \cmidrule(lr){4-5} \cmidrule(lr){6-7} \cmidrule(lr){8-9} \cmidrule(lr){10-11} \cmidrule(lr){12-13}
 & \textbf{F} & \textbf{B} & \textbf{F} & \textbf{B} & \textbf{F} & \textbf{B} & \textbf{F} & \textbf{B} & \textbf{F} & \textbf{B} & \textbf{F} & \textbf{B} \\
\midrule
MoE-1  & 91.8  & 77.1  & 12.1 & 16.2 & \textbf{8.4}  & \textbf{12.4} & 32.2  & 35.9  & 21.2 & 40.1 & \textbf{17.1} & \textbf{35.9} \\
MoE-2  & 82.0  & 63.7  & ---  & ---  & \textbf{5.4}  & \textbf{7.0}  & 29.3  & 19.2  & ---  & ---  & \textbf{10.3} & \textbf{19.1} \\
MoE-3  & 86.6  & 73.6  & 12.0 & 16.1 & \textbf{8.3}  & \textbf{12.4} & 32.1  & 36.0  & 21.1 & 40.1 & \textbf{17.0} & \textbf{36.0} \\
MoE-4  & 127.9 & 122.6 & ---  & ---  & \textbf{18.8} & \textbf{28.3} & 64.4  & 74.7  & 51.9 & 89.0 & \textbf{45.3} & 82.4  \\
MoE-5  & ---   & ---   & ---  & ---  & \textbf{9.4}  & \textbf{13.0} & 120.5 & 165.1 & ---  & ---  & \textbf{99.9} & 205.1 \\
MoE-6  & 46.6  & 38.0  & 11.7 & 14.8 & \textbf{9.1}  & \textbf{12.3} & 55.1  & 29.5  & 16.5 & 27.8 & \textbf{14.4} & \textbf{25.6} \\
MoE-7  & 75.1  & 75.4  & ---  & ---  & \textbf{20.7} & \textbf{31.7} & 81.7  & 83.9  & ---  & ---  & \textbf{46.1} & 87.3  \\
MoE-9  & 50.4  & 42.4  & 14.4 & 17.7 & \textbf{10.4} & \textbf{13.7} & 45.5  & 30.2  & 17.3 & 27.8 & \textbf{13.7} & \textbf{24.3} \\
MoE-10 & 31.6  & 16.7  & 7.7  & 7.7  & \textbf{4.6}  & \textbf{4.6}  & 24.9  & 10.5  & 7.1  & 10.3 & \textbf{4.3}  & \textbf{7.5}  \\
MoE-12 & 50.6  & 43.3  & ---  & ---  & \textbf{10.5} & \textbf{15.1} & 61.6  & 38.0  & ---  & ---  & \textbf{24.1} & 44.8  \\
\bottomrule
\end{tabular}
\end{table}

To assess scalability for long-context training, we evaluate 128k seqlen. We show the absolute latency in Table~\ref{tab:longct}.
On Cluster 1, \ours{} maintains a robust lead: {6.90$\times$} vs. Serial and {1.28$\times$} vs. COMET (forward).
On Cluster 2, the gap is {2.38$\times$} vs. Serial and {1.33$\times$} vs. COMET.
At this scale, the Serial baseline becomes more competitive because the sheer volume of computation masks the constant overheads of synchronization. However, \ours{} still delivers $>10\%$ improvement for forward+backward. In the context of large-scale training clusters with thousands of GPUs, a 10\% improvement in MFU translates to millions of dollars in saved compute time.

\subsection{End-to-End Training Throughput}
We deployed \ours{} in a production training run on 128 GPUs (16 nodes of Cluster 1) with a sequence length of 512k. \ours{} improved the training throughput from 127 billion tokens/day to {138 billion tokens/day}, which is a {1.09$\times$} speedup. Crucially, this performance gain comes with bitwise reproducibility, ensuring that the training trajectory remains mathematically identical to the verified serial baseline.


\subsection{Ablation Study}
\label{sec:ablation}

{To quantify the individual contribution of each optimization in \ours{}, we evaluate three cumulative configurations on Cluster~1 (8k sequence length), measuring the end-to-end forward and backward pass latency (ms) of the MoE layer.
1) \textbf{O}: \ours{} MegaKernel with computation-communication overlap, deterministic token ordering, and dynamic scheduling, but without bandwidth optimization or auto-tuning.
2) \textbf{B}: \textbf{O} plus the Relay Worker bandwidth optimization, which reduces NVLink traffic via intra-rank multicast.
3) \textbf{A}: \textbf{B} plus the analytical auto-tuner for optimal SM and warp allocation. Priority-based token scheduling is also included. This is the full \ours{} system.

Results are shown in Table~\ref{tab:ablation}. The bandwidth optimization (\textbf{O}$\to$\textbf{B}) reduces latency by 1.06$\times$--1.36$\times$ by replacing redundant inter-rank transmissions with intra-rank HBM copies via Relay Workers. The benefit scales with Top-$k$ and the number of experts per rank, which determines the probability of routing multiple tokens to the same destination GPU. The auto-tuner (\textbf{B}$\to$\textbf{A}) contributes an additional 1.15$\times$--1.68$\times$ speedup by identifying the optimal SM partition and warp allocation per workload, confirming that the performance model navigates the $\sim$10$^5$ configuration space effectively across diverse model architectures. The full \ours{} system (\textbf{A}) outperforms COMET by 1.24$\times$--1.73$\times$ across all supported configurations.}

\begin{table}[t]
\centering
\caption{{Ablation study on Cluster~1 for forward+backward latency (ms). O: overlap with dynamic scheduling; B: +bandwidth optimization; A: +auto-tuning with priority-based token scheduling. `-': not supported by COMET.}}
\label{tab:ablation}
\setlength{\tabcolsep}{3pt}
\begin{tabular}{lcccccccc}
\toprule
\textbf{ID} & \textbf{O} & \textbf{T} & \textbf{A} & \textbf{COMET} & \textbf{O$\to$B} & \textbf{B$\to$A} & \textbf{A vs COMET} \\
\midrule
MoE-1  &  5.25 &  4.55 &  3.47 &  5.15 & 1.15$\times$ & 1.31$\times$ & 1.48$\times$ \\
MoE-2  &  3.99 &  3.66 &  3.00 &  -  & 1.09$\times$ & 1.22$\times$ & -          \\
MoE-3  &  4.95 &  4.25 &  3.12 &  4.85 & 1.16$\times$ & 1.36$\times$ & 1.55$\times$ \\
MoE-4  &  8.92 &  7.65 &  4.67 &  7.77 & 1.16$\times$ & 1.64$\times$ & 1.66$\times$ \\
MoE-5  & 17.70 & 13.54 &  9.39 & 14.83 & 1.31$\times$ & 1.44$\times$ & 1.58$\times$ \\
MoE-6  &  6.07 &  4.92 &  3.42 &  5.90 & 1.23$\times$ & 1.44$\times$ & 1.73$\times$ \\
MoE-7  &  9.52 &  7.02 &  4.44 &  6.42 & 1.36$\times$ & 1.58$\times$ & 1.44$\times$ \\
MoE-8  & 15.63 & 12.08 &  8.56 & 11.82 & 1.29$\times$ & 1.41$\times$ & 1.38$\times$ \\
MoE-9  &  7.35 &  5.75 &  3.43 &  5.44 & 1.28$\times$ & 1.68$\times$ & 1.59$\times$ \\
MoE-10 &  4.62 &  4.37 &  3.80 &  4.70 & 1.06$\times$ & 1.15$\times$ & 1.24$\times$ \\
MoE-11 & 18.94 & 15.10 & 10.77 & 15.69 & 1.25$\times$ & 1.40$\times$ & 1.46$\times$ \\
MoE-12 &  6.76 &  5.33 &  3.37 &  -  & 1.27$\times$ & 1.58$\times$ & -          \\
\bottomrule
\end{tabular}
\end{table}

\section{Related Work and Discussion}
\label{sec:related_work}


\subsection{High-Performance Kernels}

\textbf{Expert-tuned High-performance Libraries.}
Libraries such as CuBLAS~\cite{cublas}, CuDNN~\cite{cudnn}, and CUTLASS/CuTe~\cite{cutlass} represent the state-of-the-art in compute-bound kernels. By manually managing the memory hierarchy (leveraging features like TMA and asynchronous copy) they achieve near-peak Tensor Core utilization. Specialized attention kernels like FlashAttention~\cite{flashattention} and FlashInfer~\cite{flashinfer} further optimize IO complexity.
On the communication side, NCCL~\cite{nccl} serves as the de facto standard for collective operations (AllReduce, AllGather). However, NCCL is optimized primarily for dense tensor parallelism and lacks specialized support for the sparse, dynamic patterns of EP. To address this, DeepEP~\cite{deepep} was developed, utilizing low-level NVSHMEM~\cite{nvshmem} intrinsics (IBRC, IBGDA) to initiate peer-to-peer communication directly from the GPU, bypassing the host CPU and saturating physical bandwidth.

\textbf{Tensor Compilers and DSLs.}
To lower the barrier of kernel development, compilers like Halide~\cite{halide} and TVM~\cite{tvm} introduced the separation of compute definitions from scheduling primitives. Recent advances focus on tile-centric programming models: Triton~\cite{triton} and TileLang~\cite{tilelang} expose block-level semantics that balance ease of use with performance. NVIDIA's CuTile~\cite{cutile} and CuTeDSL~\cite{cutlass} provide similar abstractions within Python. In the communication domain, MSCCL~\cite{taccl, sccl, gc3} and MSCCL++~\cite{msccl++} allow programmable synthesis of collective algorithms using primitive operations (put/get/signal), enabling topology-aware optimization. \ours{} builds upon Triton-distributed~\cite{triton-dist}, extending its capabilities to support fused computational and communication pipelines.

\subsection{Computation-Communication Overlap}

\textbf{Multi-Stream Approaches.}
Frameworks like Megatron-LM~\cite{megatron-lm}, Centauri~\cite{centauri}, and TorchTitan~\cite{torchtitan} typically employ a multi-stream execution model. Computation and communication kernels are placed on separate CUDA streams, relying on the CPU to manage synchronization.
This approach suffers from two limitations. First, splitting a large GEMM into smaller chunks to fit into overlap slots reduces the number of threadblocks per kernel. This exacerbates the {wave quantization effect}, where the tail wave of thread blocks under-utilizes the GPU, leading to degraded arithmetic intensity. Second, the reliance on CPU intervention for kernel launching and stream synchronization introduces non-negligible jitter and overhead, often destabilizing the overlap schedule at scale.

\textbf{Device-Side Signaling.}
Systems like CoCoNet~\cite{coconet}, COMET~\cite{comet}, FLUX~\cite{flux}, and TileLink~\cite{tilelink} mitigate CPU overhead by using device-side semaphores (global memory flags) to synchronize distinct compute and communication kernels. While this reduces host intervention, it still relies on dual-stream execution.
DeepSeek-V3~\cite{deepseek-v3} adopts a more aggressive strategy by overlapping the communication of the next micro-batch with the computation of the current one. While effective for specific architectures, this dual-batch approach is not suitable for FSDP training, where splitting two batches will alter the gradient accumulation order. Due to the non-associativity of floating-point arithmetic, this compromises bitwise reproducibility, which is unacceptable for rigorous training scenarios.
In contrast, \ours{} achieves overlap within a {single stream} via a fused MegaKernel. By scheduling resources at the SM level, we eliminate both CPU overhead and the need for numerical compromise.
Recent work such as Mirage~\cite{mirage} and Spector et al.~\cite{hazymega} uses MegaKernel for inference. They fuse the whole LLM into one CUDA kernel. This whole LLM fusion approach is efficient for batch 1 inference, but for large batch training, the development complexity is unaffordable and the performance gain is marginal.

\subsection{MoE Training Frameworks}
The efficiency of MoE training has been a focal point of recent research.
TransformerEngine~\cite{te} and vLLM~\cite{vllm} provide highly optimized GroupGEMM kernels. {Dynamic-centric} works like MegaBlocks \cite{megablocks} propose block-sparse formats to handle variable token loads without padding. {System-level} works such as HeterMoE~\cite{hetermoe} optimize for heterogeneous clusters, while DeepSpeed-MoE~\cite{deepspeed-moe} focuses on model compression and communication scheduling. Mega Scale-MoE \cite{moescale} integrates FLUX~\cite{flux} to hide communication costs in large-scale runs.
Despite these advancements, adoption in production-grade frameworks (e.g., Megatron-LM) remains conservative. The integration of ad-hoc optimization kernels often introduces complexity and fragility. \ours{} bridges this gap by providing a unified, mathematically rigorous abstraction that delivers the performance benefits of aggressive overlap while maintaining the stability and precision required for industrial-grade LLM training.

\subsection{MoE Inference Frameworks}
{Inference frameworks including DeepSeek-EPLB~\cite{deepseek-eplb}, FasterMoE~\cite{fastermoe}, MegaScale-Infer~\cite{megascale-infer} balances workloads for different experts by replicating expert weights to different ranks (EPLB). \ours{} is orthogonal to EPLP. \ours{} is orthogonal to expert placement strategies, the two approaches operate at different levels and can be combined to further improve inference performance.}

\subsection{Discussion}

\noindent
{\textbf{Portability to Other Hardware}
The concept of \ours{} can be generalized to other GPU platforms including AMD GPUs and latest NVIDIA GPUs. Also, Triton-distributed support AMD GPUs, which provides the possibility to migrate current implementations to AMD GPUs without major modifications.}

\noindent
{\textbf{Performance Impact of Triton}
\ours{} is implemented using Triton-distributed, which can achieve comparable performance to CUDA kernels for GroupGEMM on Hopper GPUs. While Triton-based kernels may underperform CUDA kernels on the latest GPU architectures, \ours{} design is language-agnostic and can be readily re-implemented in CUDA to close this gap, which has been proven in DeepGEMM~\cite{deepgemm} (\url{https://github.com/deepseek-ai/DeepGEMM/pull/304})

}
\section{Conclusion}
\label{sec:conclusion}

In this paper, we presented \ours{}, a unified system that mitigates the communication bottlenecks in Expert Parallelism via a composable {MegaKernel} architecture. Unlike prior works, \ours{} achieves fine-grained computation-communication overlap while strictly guaranteeing numerical consistency through deterministic token ordering. Our evaluation on NVIDIA Hopper clusters confirms that \ours{} delivers significant speedups over production baselines like COMET, offering a scalable and high-precision solution for next-generation LLM training.

\clearpage

\bibliographystyle{plainnat}
\bibliography{reference}

\begin{thebibliography}{51}
\providecommand{\natexlab}[1]{#1}
\providecommand{\url}[1]{\texttt{#1}}
\expandafter\ifx\csname urlstyle\endcsname\relax
  \providecommand{\doi}[1]{doi: #1}\else
  \providecommand{\doi}{doi: \begingroup \urlstyle{rm}\Url}\fi

\bibitem[Bai et~al.(2023)Bai, Bai, Yang, Wang, Tan, Wang, Lin, Zhou, and Zhou]{qwen-vl}
Jinze Bai, Shuai Bai, Shusheng Yang, Shijie Wang, Sinan Tan, Peng Wang, Junyang Lin, Chang Zhou, and Jingren Zhou.
\newblock Qwen-vl: {A} frontier large vision-language model with versatile abilities.
\newblock \emph{CoRR}, abs/2308.12966, 2023.
\newblock \doi{10.48550/ARXIV.2308.12966}.
\newblock URL \url{https://doi.org/10.48550/arXiv.2308.12966}.

\bibitem[Bai et~al.(2025)Bai, Bao, Chen, Chen, Chen, Chen, Chen, Chen, Chen, Chen, Cui, Ding, Dong, Du, Du, Du, Du, Fan, Feng, Fu, Gao, Gao, Gao, Gao, Gu, Guan, Guo, Guo, Hu, Hao, He, He, He, Hong, Hu, Hu, Huang, Huang, Huang, Jiang, Jiang, Jin, Kang, Lai, Li, Li, Li, Li, Li, Li, Li, Li, Li, Lin, Lin, Lin, Liu, Liu, Liu, Liu, Liu, Liu, Liu, Liu, Liu, Liu, Liu, Liu, Liu, Liu, Liu, Lu, Lu, Ma, Ma, Ma, Mao, Mei, Men, Miao, Pan, Peng, Qin, Qu, Shang, Shi, Shi, Song, Su, Su, Sun, Sung, Tang, Tao, Teng, Wang, Wang, Wang, and Wang]{kimi-k2}
Yifan Bai, Yiping Bao, Guanduo Chen, Jiahao Chen, Ningxin Chen, Ruijue Chen, Yanru Chen, Yuankun Chen, Yutian Chen, Zhuofu Chen, Jialei Cui, Hao Ding, Mengnan Dong, Angang Du, Chenzhuang Du, Dikang Du, Yulun Du, Yu~Fan, Yichen Feng, Kelin Fu, Bofei Gao, Hongcheng Gao, Peizhong Gao, Tong Gao, Xinran Gu, Longyu Guan, Haiqing Guo, Jianhang Guo, Hao Hu, Xiaoru Hao, Tianhong He, Weiran He, Wenyang He, Chao Hong, Yangyang Hu, Zhenxing Hu, Weixiao Huang, Zhiqi Huang, Zihao Huang, Tao Jiang, Zhejun Jiang, Xinyi Jin, Yongsheng Kang, Guokun Lai, Cheng Li, Fang Li, Haoyang Li, Ming Li, Wentao Li, Yanhao Li, Yiwei Li, Zhaowei Li, Zheming Li, Hongzhan Lin, Xiaohan Lin, Zongyu Lin, Chengyin Liu, Chenyu Liu, Hongzhang Liu, Jingyuan Liu, Junqi Liu, Liang Liu, Shaowei Liu, T.~Y. Liu, Tianwei Liu, Weizhou Liu, Yangyang Liu, Yibo Liu, Yiping Liu, Yue Liu, Zhengying Liu, Enzhe Lu, Lijun Lu, Shengling Ma, Xinyu Ma, Yingwei Ma, Shaoguang Mao, Jie Mei, Xin Men, Yibo Miao, Siyuan Pan, Yebo Peng, Ruoyu Qin, Bowen Qu, Zeyu Shang,
  Lidong Shi, Shengyuan Shi, Feifan Song, Jianlin Su, Zhengyuan Su, Xinjie Sun, Flood Sung, Heyi Tang, Jiawen Tao, Qifeng Teng, Chensi Wang, Dinglu Wang, Feng Wang, and Haiming Wang.
\newblock Kimi {K2:} open agentic intelligence.
\newblock \emph{CoRR}, abs/2507.20534, 2025.
\newblock \doi{10.48550/ARXIV.2507.20534}.
\newblock URL \url{https://doi.org/10.48550/arXiv.2507.20534}.

\bibitem[Cai et~al.(2021)Cai, Liu, Maleki, Musuvathi, Mytkowicz, Nelson, and Saarikivi]{sccl}
Zixian Cai, Zhengyang Liu, Saeed Maleki, Madanlal Musuvathi, Todd Mytkowicz, Jacob Nelson, and Olli Saarikivi.
\newblock Synthesizing optimal collective algorithms.
\newblock In Jaejin Lee and Erez Petrank, editors, \emph{PPoPP '21: 26th {ACM} {SIGPLAN} Symposium on Principles and Practice of Parallel Programming, Virtual Event, Republic of Korea, February 27- March 3, 2021}, pages 62--75. {ACM}, 2021.
\newblock \doi{10.1145/3437801.3441620}.
\newblock URL \url{https://doi.org/10.1145/3437801.3441620}.

\bibitem[Chang et~al.(2024)Chang, Bao, Hou, Jiang, Zheng, Zhong, Zhang, Song, Jiang, Lin, Jin, and Liu]{flux}
Li{-}Wen Chang, Wenlei Bao, Qi~Hou, Chengquan Jiang, Ningxin Zheng, Yinmin Zhong, Xuanrun Zhang, Zuquan Song, Ziheng Jiang, Haibin Lin, Xin Jin, and Xin Liu.
\newblock {FLUX:} fast software-based communication overlap on gpus through kernel fusion.
\newblock \emph{CoRR}, abs/2406.06858, 2024.
\newblock \doi{10.48550/ARXIV.2406.06858}.
\newblock URL \url{https://doi.org/10.48550/arXiv.2406.06858}.

\bibitem[Chen et~al.(2024)Chen, Li, Zhu, Duan, Sun, Zhang, and Yang]{centauri}
Chang Chen, Xiuhong Li, Qianchao Zhu, Jiangfei Duan, Peng Sun, Xingcheng Zhang, and Chao Yang.
\newblock Centauri: Enabling efficient scheduling for communication-computation overlap in large model training via communication partitioning.
\newblock In Rajiv Gupta, Nael~B. Abu{-}Ghazaleh, Madan Musuvathi, and Dan Tsafrir, editors, \emph{Proceedings of the 29th {ACM} International Conference on Architectural Support for Programming Languages and Operating Systems, Volume 3, {ASPLOS} 2024, La Jolla, CA, USA, 27 April 2024- 1 May 2024}, pages 178--191. {ACM}, 2024.
\newblock \doi{10.1145/3620666.3651379}.
\newblock URL \url{https://doi.org/10.1145/3620666.3651379}.

\bibitem[Chen et~al.(2018)Chen, Moreau, Jiang, Zheng, Yan, Shen, Cowan, Wang, Hu, Ceze, Guestrin, and Krishnamurthy]{tvm}
Tianqi Chen, Thierry Moreau, Ziheng Jiang, Lianmin Zheng, Eddie~Q. Yan, Haichen Shen, Meghan Cowan, Leyuan Wang, Yuwei Hu, Luis Ceze, Carlos Guestrin, and Arvind Krishnamurthy.
\newblock {TVM:} an automated end-to-end optimizing compiler for deep learning.
\newblock In Andrea~C. Arpaci{-}Dusseau and Geoff Voelker, editors, \emph{13th {USENIX} Symposium on Operating Systems Design and Implementation, {OSDI} 2018, Carlsbad, CA, USA, October 8-10, 2018}, pages 578--594. {USENIX} Association, 2018.

\bibitem[Cowan et~al.(2022)Cowan, Maleki, Musuvathi, Saarikivi, and Xiong]{gc3}
Meghan Cowan, Saeed Maleki, Madanlal Musuvathi, Olli Saarikivi, and Yifan Xiong.
\newblock Gc3: An optimizing compiler for gpu collective communication, 2022.
\newblock URL \url{https://arxiv.org/abs/2201.11840}.

\bibitem[Dao et~al.(2022)Dao, Fu, Ermon, Rudra, and R{\'{e}}]{flashattention}
Tri Dao, Daniel~Y. Fu, Stefano Ermon, Atri Rudra, and Christopher R{\'{e}}.
\newblock Flashattention: Fast and memory-efficient exact attention with io-awareness.
\newblock \emph{CoRR}, abs/2205.14135, 2022.
\newblock \doi{10.48550/arXiv.2205.14135}.
\newblock URL \url{https://doi.org/10.48550/arXiv.2205.14135}.

\bibitem[DeepMind(2025)]{gemini3pro}
Google DeepMind.
\newblock Gemini 3 pro: Best for complex tasks and bringing creative concepts to life, 2025.
\newblock URL \url{https://deepmind.google/models/gemini/pro/}.

\bibitem[DeepSeek-AI(2025{\natexlab{a}})]{deepgemm}
DeepSeek-AI.
\newblock Deepseek deepgemm.
\newblock \url{https://github.com/deepseek-ai/DeepGEMM}, 2025{\natexlab{a}}.

\bibitem[DeepSeek-AI(2025{\natexlab{b}})]{deepseek-eplb}
DeepSeek-AI.
\newblock {EPLB}: Expert parallelism load balancer.
\newblock \url{https://github.com/deepseek-ai/EPLB}, 2025{\natexlab{b}}.
\newblock GitHub repository.

\bibitem[DeepSeek{-}AI(2025)]{deepseek-r1}
DeepSeek{-}AI.
\newblock Deepseek-r1: Incentivizing reasoning capability in llms via reinforcement learning.
\newblock \emph{CoRR}, abs/2501.12948, 2025.
\newblock \doi{10.48550/ARXIV.2501.12948}.
\newblock URL \url{https://doi.org/10.48550/arXiv.2501.12948}.

\bibitem[DeepSeek{-}AI et~al.(2024{\natexlab{a}})DeepSeek{-}AI, Liu, Feng, Wang, Wang, Liu, Zhao, Deng, Ruan, Dai, Guo, Yang, Chen, Ji, Li, Lin, Luo, Hao, Chen, Li, Zhang, Xu, Yang, Zhang, Ding, Xin, Gao, Li, Qu, Cai, Liang, Guo, Ni, Li, Chen, Yuan, Qiu, Song, Dong, Gao, Guan, Wang, Zhang, Xu, Xia, Zhao, Zhang, Li, Wang, Zhang, Zhang, Tang, Li, Tian, Huang, Wang, Zhang, Zhu, Chen, Du, Chen, Jin, Ge, Pan, Xu, Chen, Li, Lu, Zhou, Chen, Wu, Ye, Ma, Wang, Zhou, Yu, Zhou, Zheng, Wang, Pei, Yuan, Sun, Xiao, Zeng, An, Liu, Liang, Gao, Zhang, Li, Jin, Wang, Bi, Liu, Wang, Shen, Chen, Chen, Nie, and Sun]{deepseek-v2}
DeepSeek{-}AI, Aixin Liu, Bei Feng, Bin Wang, Bingxuan Wang, Bo~Liu, Chenggang Zhao, Chengqi Deng, Chong Ruan, Damai Dai, Daya Guo, Dejian Yang, Deli Chen, Dongjie Ji, Erhang Li, Fangyun Lin, Fuli Luo, Guangbo Hao, Guanting Chen, Guowei Li, Hao Zhang, Hanwei Xu, Hao Yang, Haowei Zhang, Honghui Ding, Huajian Xin, Huazuo Gao, Hui Li, Hui Qu, J.~L. Cai, Jian Liang, Jianzhong Guo, Jiaqi Ni, Jiashi Li, Jin Chen, Jingyang Yuan, Junjie Qiu, Junxiao Song, Kai Dong, Kaige Gao, Kang Guan, Lean Wang, Lecong Zhang, Lei Xu, Leyi Xia, Liang Zhao, Liyue Zhang, Meng Li, Miaojun Wang, Mingchuan Zhang, Minghua Zhang, Minghui Tang, Mingming Li, Ning Tian, Panpan Huang, Peiyi Wang, Peng Zhang, Qihao Zhu, Qinyu Chen, Qiushi Du, R.~J. Chen, R.~L. Jin, Ruiqi Ge, Ruizhe Pan, Runxin Xu, Ruyi Chen, S.~S. Li, Shanghao Lu, Shangyan Zhou, Shanhuang Chen, Shaoqing Wu, Shengfeng Ye, Shirong Ma, Shiyu Wang, Shuang Zhou, Shuiping Yu, Shunfeng Zhou, Size Zheng, Tao Wang, Tian Pei, Tian Yuan, Tianyu Sun, W.~L. Xiao, Wangding Zeng, Wei An, Wen
  Liu, Wenfeng Liang, Wenjun Gao, Wentao Zhang, X.~Q. Li, Xiangyue Jin, Xianzu Wang, Xiao Bi, Xiaodong Liu, Xiaohan Wang, Xiaojin Shen, Xiaokang Chen, Xiaosha Chen, Xiaotao Nie, and Xiaowen Sun.
\newblock Deepseek-v2: {A} strong, economical, and efficient mixture-of-experts language model.
\newblock \emph{CoRR}, abs/2405.04434, 2024{\natexlab{a}}.
\newblock \doi{10.48550/ARXIV.2405.04434}.
\newblock URL \url{https://doi.org/10.48550/arXiv.2405.04434}.

\bibitem[DeepSeek{-}AI et~al.(2024{\natexlab{b}})DeepSeek{-}AI, Liu, Feng, Xue, Wang, Wu, Lu, Zhao, Deng, Zhang, Ruan, Dai, Guo, Yang, Chen, Ji, Li, Lin, Dai, Luo, Hao, Chen, Li, Zhang, Bao, Xu, Wang, Zhang, Ding, Xin, Gao, Li, Qu, Cai, Liang, Guo, Ni, Li, Wang, Chen, Chen, Yuan, Qiu, Li, Song, Dong, Hu, Gao, Guan, Huang, Yu, Wang, Zhang, Xu, Xia, Zhao, Wang, Zhang, Li, Wang, Zhang, Zhang, Tang, Li, Tian, Huang, Wang, Zhang, Wang, Zhu, Chen, Du, Chen, Jin, Ge, Zhang, Pan, Wang, Xu, Zhang, Chen, Li, Lu, Zhou, Chen, Wu, Ye, Ye, Ma, Wang, Zhou, Yu, Zhou, Pan, Wang, Yun, Pei, Sun, Xiao, and Zeng]{deepseek-v3}
DeepSeek{-}AI, Aixin Liu, Bei Feng, Bing Xue, Bingxuan Wang, Bochao Wu, Chengda Lu, Chenggang Zhao, Chengqi Deng, Chenyu Zhang, Chong Ruan, Damai Dai, Daya Guo, Dejian Yang, Deli Chen, Dongjie Ji, Erhang Li, Fangyun Lin, Fucong Dai, Fuli Luo, Guangbo Hao, Guanting Chen, Guowei Li, H.~Zhang, Han Bao, Hanwei Xu, Haocheng Wang, Haowei Zhang, Honghui Ding, Huajian Xin, Huazuo Gao, Hui Li, Hui Qu, J.~L. Cai, Jian Liang, Jianzhong Guo, Jiaqi Ni, Jiashi Li, Jiawei Wang, Jin Chen, Jingchang Chen, Jingyang Yuan, Junjie Qiu, Junlong Li, Junxiao Song, Kai Dong, Kai Hu, Kaige Gao, Kang Guan, Kexin Huang, Kuai Yu, Lean Wang, Lecong Zhang, Lei Xu, Leyi Xia, Liang Zhao, Litong Wang, Liyue Zhang, Meng Li, Miaojun Wang, Mingchuan Zhang, Minghua Zhang, Minghui Tang, Mingming Li, Ning Tian, Panpan Huang, Peiyi Wang, Peng Zhang, Qiancheng Wang, Qihao Zhu, Qinyu Chen, Qiushi Du, R.~J. Chen, R.~L. Jin, Ruiqi Ge, Ruisong Zhang, Ruizhe Pan, Runji Wang, Runxin Xu, Ruoyu Zhang, Ruyi Chen, S.~S. Li, Shanghao Lu, Shangyan Zhou,
  Shanhuang Chen, Shaoqing Wu, Shengfeng Ye, Shengfeng Ye, Shirong Ma, Shiyu Wang, Shuang Zhou, Shuiping Yu, Shunfeng Zhou, Shuting Pan, T.~Wang, Tao Yun, Tian Pei, Tianyu Sun, W.~L. Xiao, and Wangding Zeng.
\newblock Deepseek-v3 technical report.
\newblock \emph{CoRR}, abs/2412.19437, 2024{\natexlab{b}}.
\newblock \doi{10.48550/ARXIV.2412.19437}.
\newblock URL \url{https://doi.org/10.48550/arXiv.2412.19437}.

\bibitem[Gale et~al.(2023)Gale, Narayanan, Young, and Zaharia]{megablocks}
Trevor Gale, Deepak Narayanan, Cliff Young, and Matei Zaharia.
\newblock Megablocks: Efficient sparse training with mixture-of-experts.
\newblock In Dawn Song, Michael Carbin, and Tianqi Chen, editors, \emph{Proceedings of the Sixth Conference on Machine Learning and Systems, MLSys 2023, Miami, FL, USA, June 4-8, 2023}. mlsys.org, 2023.
\newblock URL \url{https://proceedings.mlsys.org/paper\_files/paper/2023/hash/5a54f79333768effe7e8927bcccffe40-Abstract-mlsys2023.html}.

\bibitem[He et~al.(2022)He, Zhai, Antunes, Wang, Luo, Shi, and Li]{fastermoe}
Jiaao He, Jidong Zhai, Tiago Antunes, Haojie Wang, Fuwen Luo, Shangfeng Shi, and Qin Li.
\newblock Fastermoe: modeling and optimizing training of large-scale dynamic pre-trained models.
\newblock In \emph{Proceedings of the 27th ACM SIGPLAN Symposium on Principles and Practice of Parallel Programming}, pages 120--134, 2022.

\bibitem[Jangda et~al.(2022)Jangda, Huang, Liu, Sabet, Maleki, Miao, Musuvathi, Mytkowicz, and Saarikivi]{coconet}
Abhinav Jangda, Jun Huang, Guodong Liu, Amir Hossein~Nodehi Sabet, Saeed Maleki, Youshan Miao, Madanlal Musuvathi, Todd Mytkowicz, and Olli Saarikivi.
\newblock Breaking the computation and communication abstraction barrier in distributed machine learning workloads.
\newblock In Babak Falsafi, Michael Ferdman, Shan Lu, and Thomas~F. Wenisch, editors, \emph{{ASPLOS} '22: 27th {ACM} International Conference on Architectural Support for Programming Languages and Operating Systems, Lausanne, Switzerland, 28 February 2022 - 4 March 2022}, pages 402--416. {ACM}, 2022.
\newblock \doi{10.1145/3503222.3507778}.
\newblock URL \url{https://doi.org/10.1145/3503222.3507778}.

\bibitem[Jin et~al.(2025)Jin, Jiang, Bai, Zhong, Liu, Li, Zheng, Wang, Xie, Huang, Heng, Ma, Bao, Zheng, Peng, Lin, Liu, Jin, and Liu]{moescale}
Chao Jin, Ziheng Jiang, Zhihao Bai, Zheng Zhong, Juncai Liu, Xiang Li, Ningxin Zheng, Xi~Wang, Cong Xie, Qi~Huang, Wen Heng, Yiyuan Ma, Wenlei Bao, Size Zheng, Yanghua Peng, Haibin Lin, Xuanzhe Liu, Xin Jin, and Xin Liu.
\newblock Megascale-moe: Large-scale communication-efficient training of mixture-of-experts models in production.
\newblock \emph{CoRR}, abs/2505.11432, 2025.
\newblock \doi{10.48550/ARXIV.2505.11432}.
\newblock URL \url{https://doi.org/10.48550/arXiv.2505.11432}.

\bibitem[Kwon et~al.(2023)Kwon, Li, Zhuang, Sheng, Zheng, Yu, Gonzalez, Zhang, and Stoica]{vllm}
Woosuk Kwon, Zhuohan Li, Siyuan Zhuang, Ying Sheng, Lianmin Zheng, Cody~Hao Yu, Joseph Gonzalez, Hao Zhang, and Ion Stoica.
\newblock Efficient memory management for large language model serving with pagedattention.
\newblock In Jason Flinn, Margo~I. Seltzer, Peter Druschel, Antoine Kaufmann, and Jonathan Mace, editors, \emph{Proceedings of the 29th Symposium on Operating Systems Principles, {SOSP} 2023, Koblenz, Germany, October 23-26, 2023}, pages 611--626. {ACM}, 2023.
\newblock \doi{10.1145/3600006.3613165}.
\newblock URL \url{https://doi.org/10.1145/3600006.3613165}.

\bibitem[Liang et~al.(2024)Liang, Liu, Wright, Constable, Gu, Huang, Zhang, Feng, Huang, Wang, Purandare, Nadathur, and Idreos]{torchtitan}
Wanchao Liang, Tianyu Liu, Less Wright, Will Constable, Andrew Gu, Chien-Chin Huang, Iris Zhang, Wei Feng, Howard Huang, Junjie Wang, Sanket Purandare, Gokul Nadathur, and Stratos Idreos.
\newblock Torchtitan: One-stop pytorch native solution for production ready llm pre-training, 2024.
\newblock URL \url{https://arxiv.org/abs/2410.06511}.

\bibitem[Liu et~al.(2025)Liu, Wang, Fu, Miao, Zhu, Nie, and Cui]{netmoe}
Xinyi Liu, Yujie Wang, Fangcheng Fu, Xupeng Miao, Shenhan Zhu, Xiaonan Nie, and Bin Cui.
\newblock Netmoe: Accelerating moe training through dynamic sample placement.
\newblock In \emph{The Thirteenth International Conference on Learning Representations, {ICLR} 2025, Singapore, April 24-28, 2025}. OpenReview.net, 2025.
\newblock URL \url{https://openreview.net/forum?id=1qP3lsatCR}.

\bibitem[Lu et~al.(2024)Lu, Liu, Zhang, Wang, Dong, Liu, Sun, Ren, Li, Yang, Sun, Deng, Xu, Xie, and Ruan]{deepseek-vl}
Haoyu Lu, Wen Liu, Bo~Zhang, Bingxuan Wang, Kai Dong, Bo~Liu, Jingxiang Sun, Tongzheng Ren, Zhuoshu Li, Hao Yang, Yaofeng Sun, Chengqi Deng, Hanwei Xu, Zhenda Xie, and Chong Ruan.
\newblock Deepseek-vl: Towards real-world vision-language understanding.
\newblock \emph{CoRR}, abs/2403.05525, 2024.
\newblock \doi{10.48550/ARXIV.2403.05525}.
\newblock URL \url{https://doi.org/10.48550/arXiv.2403.05525}.

\bibitem[Narayanan et~al.(2021)Narayanan, Shoeybi, Casper, LeGresley, Patwary, Korthikanti, Vainbrand, Kashinkunti, Bernauer, Catanzaro, Phanishayee, and Zaharia]{megatron-lm}
Deepak Narayanan, Mohammad Shoeybi, Jared Casper, Patrick LeGresley, Mostofa Patwary, Vijay Korthikanti, Dmitri Vainbrand, Prethvi Kashinkunti, Julie Bernauer, Bryan Catanzaro, Amar Phanishayee, and Matei Zaharia.
\newblock Efficient large-scale language model training on {GPU} clusters using megatron-lm.
\newblock In Bronis~R. de~Supinski, Mary~W. Hall, and Todd Gamblin, editors, \emph{International Conference for High Performance Computing, Networking, Storage and Analysis, {SC} 2021, St. Louis, Missouri, USA, November 14-19, 2021}, page~58. {ACM}, 2021.
\newblock \doi{10.1145/3458817.3476209}.
\newblock URL \url{https://doi.org/10.1145/3458817.3476209}.

\bibitem[NVIDIA(2022)]{cublas}
NVIDIA.
\newblock {cuBLAS}, 2022.
\newblock URL \url{https://developer.nvidia.com/cublas}.

\bibitem[Nvidia(2022)]{cutlass}
Nvidia.
\newblock Cutlass, 2022.
\newblock URL \url{https://github.com/NVIDIA/cutlass}.

\bibitem[NVIDIA(2022)]{te}
NVIDIA.
\newblock {Transformer Engine}, 2022.
\newblock URL \url{https://github.com/NVIDIA/TransformerEngine}.

\bibitem[NVIDIA(2024)]{nccl}
NVIDIA.
\newblock Nvidia collective communications library.
\newblock \url{https://developer.nvidia.com/nccl}, 2024.

\bibitem[NVIDIA(2025)]{nvshmem}
NVIDIA.
\newblock {NVSHMEM}, 2025.
\newblock URL \url{https://docs.nvidia.com/nvshmem/api/using.html}.

\bibitem[Nvidia(2026{\natexlab{a}})]{cudnn}
Nvidia.
\newblock Cudnn, 2026{\natexlab{a}}.
\newblock URL \url{https://developer.nvidia.com/cudnn}.

\bibitem[Nvidia(2026{\natexlab{b}})]{cutile}
Nvidia.
\newblock Cutile, 2026{\natexlab{b}}.
\newblock URL \url{https://github.com/NVIDIA/cutile-python}.

\bibitem[Ragan{-}Kelley et~al.(2013)Ragan{-}Kelley, Barnes, Adams, Paris, Durand, and Amarasinghe]{halide}
Jonathan Ragan{-}Kelley, Connelly Barnes, Andrew Adams, Sylvain Paris, Fr{\'{e}}do Durand, and Saman~P. Amarasinghe.
\newblock Halide: a language and compiler for optimizing parallelism, locality, and recomputation in image processing pipelines.
\newblock In \emph{{ACM} {SIGPLAN} Conference on Programming Language Design and Implementation, {PLDI} '13, Seattle, WA, USA, June 16-19, 2013}, pages 519--530, 2013.
\newblock \doi{10.1145/2491956.2462176}.
\newblock URL \url{https://doi.org/10.1145/2491956.2462176}.

\bibitem[Rajbhandari et~al.(2022)Rajbhandari, Li, Yao, Zhang, Aminabadi, Awan, Rasley, and He]{deepspeed-moe}
Samyam Rajbhandari, Conglong Li, Zhewei Yao, Minjia Zhang, Reza~Yazdani Aminabadi, Ammar~Ahmad Awan, Jeff Rasley, and Yuxiong He.
\newblock Deepspeed-moe: Advancing mixture-of-experts inference and training to power next-generation {AI} scale.
\newblock In Kamalika Chaudhuri, Stefanie Jegelka, Le~Song, Csaba Szepesv{\'{a}}ri, Gang Niu, and Sivan Sabato, editors, \emph{International Conference on Machine Learning, {ICML} 2022, 17-23 July 2022, Baltimore, Maryland, {USA}}, volume 162 of \emph{Proceedings of Machine Learning Research}, pages 18332--18346. {PMLR}, 2022.
\newblock URL \url{https://proceedings.mlr.press/v162/rajbhandari22a.html}.

\bibitem[Shah et~al.(2023)Shah, Chidambaram, Cowan, Maleki, Musuvathi, Mytkowicz, Nelson, and Saarikivi]{taccl}
Aashaka Shah, Vijay Chidambaram, Meghan Cowan, Saeed Maleki, Madan Musuvathi, Todd Mytkowicz, Jacob Nelson, and Olli Saarikivi.
\newblock {TACCL:} guiding collective algorithm synthesis using communication sketches.
\newblock In Mahesh Balakrishnan and Manya Ghobadi, editors, \emph{20th {USENIX} Symposium on Networked Systems Design and Implementation, {NSDI} 2023, Boston, MA, April 17-19, 2023}, pages 593--612. {USENIX} Association, 2023.
\newblock URL \url{https://www.usenix.org/conference/nsdi23/presentation/shah}.

\bibitem[Shah et~al.(2025)Shah, Jangda, Li, Rocha, Hwang, Jose, Musuvathi, Saarikivi, Cheng, Zhou, Dathathri, Maleki, and Yang]{msccl++}
Aashaka Shah, Abhinav Jangda, Binyang Li, Caio Rocha, Changho Hwang, Jithin Jose, Madan Musuvathi, Olli Saarikivi, Peng Cheng, Qinghua Zhou, Roshan Dathathri, Saeed Maleki, and Ziyue Yang.
\newblock Msccl++: Rethinking gpu communication abstractions for cutting-edge ai applications, 2025.
\newblock URL \url{https://arxiv.org/abs/2504.09014}.

\bibitem[Singh et~al.(2025)Singh, Fry, Perelman, Tart, Ganesh, El-Kishky, McLaughlin, Low, Ostrow, Ananthram, Nathan, Luo, Helyar, Madry, Efremov, Spyra, Baker-Whitcomb, Beutel, Karpenko, Makelov, Neitz, Wei, Barr, Kirchmeyer, Ivanov, Christakis, Gillespie, Tam, Bennett, Wan, Huang, Sandjideh, Yang, Kumar, Saraiva, Vallone, Gheorghe, Garcia, Braunstein, Liu, Schmidt, Mereskin, Mishchenko, Applebaum, Rogerson, Rajan, Wei, Kotha, Srivastava, Agrawal, Vijayvergiya, Tyra, Nair, Nayak, Eggers, Ji, Hoover, Chen, Chen, Barak, Minaiev, Hao, Baker, Lightcap, McKinzie, Wang, Quinn, Fioca, Hsu, Yang, Yu, Zhang, Brenner, Zetino, Raymond, Lugaresi, Paz, Hudson, Whitney, Li, Chen, Cole, Voss, Ding, Shen, Huang, Colby, Hallacy, Koch, Lu, Kaplan, Kim, Minott-Henriques, Frey, Yu, Czarnecki, Reid, Wei, Decareaux, Scheau, Zhang, Forbes, Tang, Goldberg, Roberts, Palmie, Kappler, Levine, Wright, Leo, Lin, Robinson, Grabb, Chen, Lim, Salama, Bhattacharjee, Tsipras, Li, Yu, Strouse, Williams, Hunn, Bayes, Arbus, Akyurek, Le,
  Widmann, Yani, Proehl, Sert, Cheung, Schwartz, Han, Jiang, Mitchell, Sigler, Wallace, Ritter, Kavanaugh, Mays, Nikishin, Li, Such, de~Avila Belbute~Peres, Raso, Bekerman, Tsimpourlas, Chantzis, Song, Zhang, Raila, McGrath, Briggs, Yang, Parascandolo, Chabot, Kim, Zhao, Valiant, Leclerc, Salman, Wang, Sheng, Jiang, Wang, Jin, Sikchi, Schmidt, Aspegren, Chen, Qiu, Lightman, Covert, Kivlichan, Silber, Sohl, Hammoud, Clavera, Lan, Akkaya, Kostrikov, Kofman, Etinger, Singal, Hehir, Huh, Pan, Wilczynski, Pachocki, Lee, Quinn, Kiros, Kalra, Samaroo, Wang, Wolfe, Chen, Wang, Harb, Han, Wang, Zhao, Chen, Yang, Tworek, Chand, Landon, Liang, Lin, Liu, Wang, Tang, Yin, Jang, Morris, Flynn, Ferstad, Heidecke, Fishbein, Hallman, Grant, Chien, Gordon, Park, Liss, Kraaijeveld, Guay, Mo, Lawson, McGrath, Vendrow, Jiao, Lee, Steele, Wang, Mao, Chen, Hayashi, Xiao, Salahi, Wu, Sekhri, Sharma, Singhal, Li, Nguyen, Gu-Lemberg, King, Liu, Stone, Yu, Ying, Georgiev, Lim, Tirumala, Miller, Ahmad, Lv, Clare, Fauconnet, Itow, Yang,
  Romaniuk, Anise, Byron, Pathak, Maksin, Lo, Ho, Jing, Wu, Xiong, Mamitsuka, Yang, McCallum, Held, Bourgeois, Engstrom, Kuhn, Feuvrier, Zhang, Switzer, Kondraciuk, Kaiser, Joglekar, Singh, Shah, Stratta, Williams, Chen, Sun, Cayton, Li, Zhang, Aljubeh, Nichols, Haines, Schwarzer, Gupta, Shah, Huang, Dong, Wang, Glaese, Carroll, Lampe, Malek, Sharman, Zhang, Wang, Pokrass, Florian, Pavlov, Wang, Chen, Wang, Feng, Bavarian, Lin, Abdool, Rohaninejad, Soto, Staudacher, LaFontaine, Marwell, Liu, Preston, Turley, Ansman, Blades, Pancha, Mikhaylin, Felix, Handa, Rai, Keskar, Brown, Nachum, Boiko, Murk, Watkins, Gleeson, Mishkin, Lesiewicz, Baltescu, Belov, Zhokhov, Pronin, Guo, Thacker, Liu, Yuan, Liu, Dias, Puckett, Arora, Mullapudi, Gaon, Miyara, Song, Aggarwal, Marsan, Yemiru, Xiong, Kshirsagar, Nuttall, Tsiupa, Eldan, Wang, James, Ziv, Shu, Nigmatullin, Jain, Talaie, Altman, Arnesen, Toizer, Toyer, Miserendino, Agarwal, Yoo, Heon, Ethersmith, Grove, Taylor, Bubeck, Banesiu, Amdo, Zhao, Wu, Santurkar, Zhao,
  Chaudhuri, Krishnaswamy, Shuaiqi, Xia, Cheng, Anadkat, Fishman, Tobin, Fu, Jain, Mei, Egoian, Kim, Golden, Mah, Lin, Imm, Sharpe, Yadlowsky, Choudhry, Eum, Sanjeev, Khan, Stramer, Wang, Xin, Gogineni, Christianson, Sanders, Patwardhan, Degry, Shadwell, Fu, Gao, Garipov, Sriskandarajah, Sherbakov, Kaftan, Hiratsuka, Wang, Song, Zhao, Peterson, Kharitonov, Chernova, Kosaraju, Kuo, Pong, Verma, Petrov, Jiang, Zhang, Zhou, Xie, Zhan, McCabe, DePue, Ellsworth, Bain, Thompson, Chen, Qi, Xiang, Shi, Dubois, Yu, Khakbaz, Wu, Qian, Lee, Chen, Zhang, Xiong, Tian, Cha, Bai, Yang, Yuan, Li, Zhang, Yang, Jin, Jiang, Wang, Wang, Liu, Stubenvoll, Dou, Wu, and Wang]{gpt5}
Aaditya Singh, Adam Fry, Adam Perelman, Adam Tart, Adi Ganesh, Ahmed El-Kishky, Aidan McLaughlin, Aiden Low, AJ~Ostrow, Akhila Ananthram, Akshay Nathan, Alan Luo, Alec Helyar, Aleksander Madry, Aleksandr Efremov, Aleksandra Spyra, Alex Baker-Whitcomb, Alex Beutel, Alex Karpenko, Alex Makelov, Alex Neitz, Alex Wei, Alexandra Barr, Alexandre Kirchmeyer, Alexey Ivanov, Alexi Christakis, Alistair Gillespie, Allison Tam, Ally Bennett, Alvin Wan, Alyssa Huang, Amy~McDonald Sandjideh, Amy Yang, Ananya Kumar, Andre Saraiva, Andrea Vallone, Andrei Gheorghe, Andres~Garcia Garcia, Andrew Braunstein, Andrew Liu, Andrew Schmidt, Andrey Mereskin, Andrey Mishchenko, Andy Applebaum, Andy Rogerson, Ann Rajan, Annie Wei, Anoop Kotha, Anubha Srivastava, Anushree Agrawal, Arun Vijayvergiya, Ashley Tyra, Ashvin Nair, Avi Nayak, Ben Eggers, Bessie Ji, Beth Hoover, Bill Chen, Blair Chen, Boaz Barak, Borys Minaiev, Botao Hao, Bowen Baker, Brad Lightcap, Brandon McKinzie, Brandon Wang, Brendan Quinn, Brian Fioca, Brian Hsu, Brian
  Yang, Brian Yu, Brian Zhang, Brittany Brenner, Callie~Riggins Zetino, Cameron Raymond, Camillo Lugaresi, Carolina Paz, Cary Hudson, Cedric Whitney, Chak Li, Charles Chen, Charlotte Cole, Chelsea Voss, Chen Ding, Chen Shen, Chengdu Huang, Chris Colby, Chris Hallacy, Chris Koch, Chris Lu, Christina Kaplan, Christina Kim, CJ~Minott-Henriques, Cliff Frey, Cody Yu, Coley Czarnecki, Colin Reid, Colin Wei, Cory Decareaux, Cristina Scheau, Cyril Zhang, Cyrus Forbes, Da~Tang, Dakota Goldberg, Dan Roberts, Dana Palmie, Daniel Kappler, Daniel Levine, Daniel Wright, Dave Leo, David Lin, David Robinson, Declan Grabb, Derek Chen, Derek Lim, Derek Salama, Dibya Bhattacharjee, Dimitris Tsipras, Dinghua Li, Dingli Yu, DJ~Strouse, Drew Williams, Dylan Hunn, Ed~Bayes, Edwin Arbus, Ekin Akyurek, Elaine~Ya Le, Elana Widmann, Eli Yani, Elizabeth Proehl, Enis Sert, Enoch Cheung, Eri Schwartz, Eric Han, Eric Jiang, Eric Mitchell, Eric Sigler, Eric Wallace, Erik Ritter, Erin Kavanaugh, Evan Mays, Evgenii Nikishin, Fangyuan Li,
  Felipe~Petroski Such, Filipe de~Avila Belbute~Peres, Filippo Raso, Florent Bekerman, Foivos Tsimpourlas, Fotis Chantzis, Francis Song, Francis Zhang, Gaby Raila, Garrett McGrath, Gary Briggs, Gary Yang, Giambattista Parascandolo, Gildas Chabot, Grace Kim, Grace Zhao, Gregory Valiant, Guillaume Leclerc, Hadi Salman, Hanson Wang, Hao Sheng, Haoming Jiang, Haoyu Wang, Haozhun Jin, Harshit Sikchi, Heather Schmidt, Henry Aspegren, Honglin Chen, Huida Qiu, Hunter Lightman, Ian Covert, Ian Kivlichan, Ian Silber, Ian Sohl, Ibrahim Hammoud, Ignasi Clavera, Ikai Lan, Ilge Akkaya, Ilya Kostrikov, Irina Kofman, Isak Etinger, Ishaan Singal, Jackie Hehir, Jacob Huh, Jacqueline Pan, Jake Wilczynski, Jakub Pachocki, James Lee, James Quinn, Jamie Kiros, Janvi Kalra, Jasmyn Samaroo, Jason Wang, Jason Wolfe, Jay Chen, Jay Wang, Jean Harb, Jeffrey Han, Jeffrey Wang, Jennifer Zhao, Jeremy Chen, Jerene Yang, Jerry Tworek, Jesse Chand, Jessica Landon, Jessica Liang, Ji~Lin, Jiancheng Liu, Jianfeng Wang, Jie Tang, Jihan Yin,
  Joanne Jang, Joel Morris, Joey Flynn, Johannes Ferstad, Johannes Heidecke, John Fishbein, John Hallman, Jonah Grant, Jonathan Chien, Jonathan Gordon, Jongsoo Park, Jordan Liss, Jos Kraaijeveld, Joseph Guay, Joseph Mo, Josh Lawson, Josh McGrath, Joshua Vendrow, Joy Jiao, Julian Lee, Julie Steele, Julie Wang, Junhua Mao, Kai Chen, Kai Hayashi, Kai Xiao, Kamyar Salahi, Kan Wu, Karan Sekhri, Karan Sharma, Karan Singhal, Karen Li, Kenny Nguyen, Keren Gu-Lemberg, Kevin King, Kevin Liu, Kevin Stone, Kevin Yu, Kristen Ying, Kristian Georgiev, Kristie Lim, Kushal Tirumala, Kyle Miller, Lama Ahmad, Larry Lv, Laura Clare, Laurance Fauconnet, Lauren Itow, Lauren Yang, Laurentia Romaniuk, Leah Anise, Lee Byron, Leher Pathak, Leon Maksin, Leyan Lo, Leyton Ho, Li~Jing, Liang Wu, Liang Xiong, Lien Mamitsuka, Lin Yang, Lindsay McCallum, Lindsey Held, Liz Bourgeois, Logan Engstrom, Lorenz Kuhn, Louis Feuvrier, Lu~Zhang, Lucas Switzer, Lukas Kondraciuk, Lukasz Kaiser, Manas Joglekar, Mandeep Singh, Mandip Shah, Manuka
  Stratta, Marcus Williams, Mark Chen, Mark Sun, Marselus Cayton, Martin Li, Marvin Zhang, Marwan Aljubeh, Matt Nichols, Matthew Haines, Max Schwarzer, Mayank Gupta, Meghan Shah, Melody Huang, Meng Dong, Mengqing Wang, Mia Glaese, Micah Carroll, Michael Lampe, Michael Malek, Michael Sharman, Michael Zhang, Michele Wang, Michelle Pokrass, Mihai Florian, Mikhail Pavlov, Miles Wang, Ming Chen, Mingxuan Wang, Minnia Feng, Mo~Bavarian, Molly Lin, Moose Abdool, Mostafa Rohaninejad, Nacho Soto, Natalie Staudacher, Natan LaFontaine, Nathan Marwell, Nelson Liu, Nick Preston, Nick Turley, Nicklas Ansman, Nicole Blades, Nikil Pancha, Nikita Mikhaylin, Niko Felix, Nikunj Handa, Nishant Rai, Nitish Keskar, Noam Brown, Ofir Nachum, Oleg Boiko, Oleg Murk, Olivia Watkins, Oona Gleeson, Pamela Mishkin, Patryk Lesiewicz, Paul Baltescu, Pavel Belov, Peter Zhokhov, Philip Pronin, Phillip Guo, Phoebe Thacker, Qi~Liu, Qiming Yuan, Qinghua Liu, Rachel Dias, Rachel Puckett, Rahul Arora, Ravi~Teja Mullapudi, Raz Gaon, Reah Miyara,
  Rennie Song, Rishabh Aggarwal, RJ~Marsan, Robel Yemiru, Robert Xiong, Rohan Kshirsagar, Rohan Nuttall, Roman Tsiupa, Ronen Eldan, Rose Wang, Roshan James, Roy Ziv, Rui Shu, Ruslan Nigmatullin, Saachi Jain, Saam Talaie, Sam Altman, Sam Arnesen, Sam Toizer, Sam Toyer, Samuel Miserendino, Sandhini Agarwal, Sarah Yoo, Savannah Heon, Scott Ethersmith, Sean Grove, Sean Taylor, Sebastien Bubeck, Sever Banesiu, Shaokyi Amdo, Shengjia Zhao, Sherwin Wu, Shibani Santurkar, Shiyu Zhao, Shraman~Ray Chaudhuri, Shreyas Krishnaswamy, Shuaiqi, Xia, Shuyang Cheng, Shyamal Anadkat, Simón~Posada Fishman, Simon Tobin, Siyuan Fu, Somay Jain, Song Mei, Sonya Egoian, Spencer Kim, Spug Golden, SQ~Mah, Steph Lin, Stephen Imm, Steve Sharpe, Steve Yadlowsky, Sulman Choudhry, Sungwon Eum, Suvansh Sanjeev, Tabarak Khan, Tal Stramer, Tao Wang, Tao Xin, Tarun Gogineni, Taya Christianson, Ted Sanders, Tejal Patwardhan, Thomas Degry, Thomas Shadwell, Tianfu Fu, Tianshi Gao, Timur Garipov, Tina Sriskandarajah, Toki Sherbakov, Tomer Kaftan,
  Tomo Hiratsuka, Tongzhou Wang, Tony Song, Tony Zhao, Troy Peterson, Val Kharitonov, Victoria Chernova, Vineet Kosaraju, Vishal Kuo, Vitchyr Pong, Vivek Verma, Vlad Petrov, Wanning Jiang, Weixing Zhang, Wenda Zhou, Wenlei Xie, Wenting Zhan, Wes McCabe, Will DePue, Will Ellsworth, Wulfie Bain, Wyatt Thompson, Xiangning Chen, Xiangyu Qi, Xin Xiang, Xinwei Shi, Yann Dubois, Yaodong Yu, Yara Khakbaz, Yifan Wu, Yilei Qian, Yin~Tat Lee, Yinbo Chen, Yizhen Zhang, Yizhong Xiong, Yonglong Tian, Young Cha, Yu~Bai, Yu~Yang, Yuan Yuan, Yuanzhi Li, Yufeng Zhang, Yuguang Yang, Yujia Jin, Yun Jiang, Yunyun Wang, Yushi Wang, Yutian Liu, Zach Stubenvoll, Zehao Dou, Zheng Wu, and Zhigang Wang.
\newblock Openai gpt-5 system card, 2025.
\newblock URL \url{https://arxiv.org/abs/2601.03267}.

\bibitem[Spector et~al.(2025)Spector, Juravsky, Sul, Dugan, Lim, Fu, Arora, and R{\'e}]{hazymega}
Benjamin Spector, Jordan Juravsky, Stuart Sul, Owen Dugan, Dylan Lim, Dan Fu, Simran Arora, and Chris R{\'e}.
\newblock Look ma, no bubbles! designing a low-latency megakernel for {LLAMA}-1{B}.
\newblock \url{https://hazyresearch.stanford.edu/blog/2025-05-27-no-bubbles}, 2025.
\newblock Hazy Research Blog.

\bibitem[TileLang-Team(2025)]{tilelang}
TileLang-Team.
\newblock Tilelang, 2025.
\newblock URL \url{https://github.com/tile-ai/tilelang}.

\bibitem[Tillet et~al.(2019)Tillet, Kung, and Cox]{triton}
Philippe Tillet, Hsiang{-}Tsung Kung, and David~D. Cox.
\newblock Triton: an intermediate language and compiler for tiled neural network computations.
\newblock In Tim Mattson, Abdullah Muzahid, and Armando Solar{-}Lezama, editors, \emph{Proceedings of the 3rd {ACM} {SIGPLAN} International Workshop on Machine Learning and Programming Languages, MAPL@PLDI 2019, Phoenix, AZ, USA, June 22, 2019}, pages 10--19. {ACM}, 2019.
\newblock \doi{10.1145/3315508.3329973}.
\newblock URL \url{https://doi.org/10.1145/3315508.3329973}.

\bibitem[Wei et~al.(2026)Wei, Sun, and Li]{deepseek-ocr2}
Haoran Wei, Yaofeng Sun, and Yukun Li.
\newblock Deepseek-ocr 2: Visual causal flow, 2026.
\newblock URL \url{https://arxiv.org/abs/2601.20552}.

\bibitem[Wu et~al.(2025{\natexlab{a}})Wu, Cheng, Liu, Shi, Ji, Ao, Velliengiri, Miao, Padon, and Jia]{mirage}
Mengdi Wu, Xinhao Cheng, Shengyu Liu, Chunan Shi, Jianan Ji, Kit Ao, Praveen Velliengiri, Xupeng Miao, Oded Padon, and Zhihao Jia.
\newblock Mirage: A multi-level superoptimizer for tensor programs.
\newblock In \emph{19th USENIX Symposium on Operating Systems Design and Implementation (OSDI 25)}, Boston, MA, July 2025{\natexlab{a}}. USENIX Association.

\bibitem[Wu et~al.(2025{\natexlab{b}})Wu, Liu, Jin, Xu, Qian, Mao, Lentz, Zhuo, and Stoica]{hetermoe}
Yongji Wu, Xueshen Liu, Shuowei Jin, Ceyu Xu, Feng Qian, Z.~Morley Mao, Matthew Lentz, Danyang Zhuo, and Ion Stoica.
\newblock Hetermoe: Efficient training of mixture-of-experts models on heterogeneous gpus.
\newblock \emph{CoRR}, abs/2504.03871, 2025{\natexlab{b}}.
\newblock \doi{10.48550/ARXIV.2504.03871}.
\newblock URL \url{https://doi.org/10.48550/arXiv.2504.03871}.

\bibitem[Yang et~al.(2025{\natexlab{a}})Yang, Li, Yang, Zhang, Hui, Zheng, Yu, Gao, Huang, Lv, Zheng, Liu, Zhou, Huang, Hu, Ge, Wei, Lin, Tang, Yang, Tu, Zhang, Yang, Yang, Zhou, Zhou, Lin, Dang, Bao, Yang, Yu, Deng, Li, Xue, Li, Zhang, Wang, Zhu, Men, Gao, Liu, Luo, Li, Tang, Yin, Ren, Wang, Zhang, Ren, Fan, Su, Zhang, Zhang, Wan, Liu, Wang, Cui, Zhang, Zhou, and Qiu]{qwen3}
An~Yang, Anfeng Li, Baosong Yang, Beichen Zhang, Binyuan Hui, Bo~Zheng, Bowen Yu, Chang Gao, Chengen Huang, Chenxu Lv, Chujie Zheng, Dayiheng Liu, Fan Zhou, Fei Huang, Feng Hu, Hao Ge, Haoran Wei, Huan Lin, Jialong Tang, Jian Yang, Jianhong Tu, Jianwei Zhang, Jian Yang, Jiaxi Yang, Jingren Zhou, Jingren Zhou, Junyang Lin, Kai Dang, Keqin Bao, Kexin Yang, Le~Yu, Lianghao Deng, Mei Li, Mingfeng Xue, Mingze Li, Pei Zhang, Peng Wang, Qin Zhu, Rui Men, Ruize Gao, Shixuan Liu, Shuang Luo, Tianhao Li, Tianyi Tang, Wenbiao Yin, Xingzhang Ren, Xinyu Wang, Xinyu Zhang, Xuancheng Ren, Yang Fan, Yang Su, Yichang Zhang, Yinger Zhang, Yu~Wan, Yuqiong Liu, Zekun Wang, Zeyu Cui, Zhenru Zhang, Zhipeng Zhou, and Zihan Qiu.
\newblock Qwen3 technical report.
\newblock \emph{CoRR}, abs/2505.09388, 2025{\natexlab{a}}.
\newblock \doi{10.48550/ARXIV.2505.09388}.
\newblock URL \url{https://doi.org/10.48550/arXiv.2505.09388}.

\bibitem[Yang et~al.(2025{\natexlab{b}})Yang, Huang, Wu, Li, Pan, Zheng, Xia, Li, and Wang]{hybrid-ep}
Weihao Yang, Hao Huang, Donglei Wu, Ningke Li, Yanqi Pan, Qiyang Zheng, Wen Xia, Shiyi Li, and Qiang Wang.
\newblock Hybridep: Scaling expert parallelism to cross-datacenter scenario via hybrid expert/data transmission.
\newblock \emph{CoRR}, abs/2510.19470, 2025{\natexlab{b}}.
\newblock \doi{10.48550/ARXIV.2510.19470}.
\newblock URL \url{https://doi.org/10.48550/arXiv.2510.19470}.

\bibitem[Ye et~al.(2025)Ye, Chen, Lai, Lin, Zhang, Wang, Chen, Kasikci, Grover, Krishnamurthy, and Ceze]{flashinfer}
Zihao Ye, Lequn Chen, Ruihang Lai, Wuwei Lin, Yineng Zhang, Stephanie Wang, Tianqi Chen, Baris Kasikci, Vinod Grover, Arvind Krishnamurthy, and Luis Ceze.
\newblock Flashinfer: Efficient and customizable attention engine for llm inference serving.
\newblock \emph{arXiv preprint arXiv:2501.01005}, 2025.
\newblock URL \url{https://arxiv.org/abs/2501.01005}.

\bibitem[Zhang et~al.(2026)Zhang, Liu, Yan, Deng, Cao, Yang, Ni, Xue, and Li]{moeblaze}
Jiyuan Zhang, Yining Liu, Siqi Yan, Lisen Deng, Jennifer Cao, Shuqi Yang, Min Ni, Bi~Xue, and Shen Li.
\newblock Moeblaze: Breaking the memory wall for efficient moe training on modern gpus, 2026.
\newblock URL \url{https://arxiv.org/abs/2601.05296}.

\bibitem[Zhang et~al.(2025{\natexlab{a}})Zhang, Zheng, Lin, Jiang, Bao, Jiang, Hou, Cui, Zheng, Chang, Chen, and Liu]{comet}
Shulai Zhang, Ningxin Zheng, Haibin Lin, Ziheng Jiang, Wenlei Bao, Chengquan Jiang, Qi~Hou, Weihao Cui, Size Zheng, Li{-}Wen Chang, Quan Chen, and Xin Liu.
\newblock Comet: Fine-grained computation-communication overlapping for mixture-of-experts.
\newblock \emph{CoRR}, abs/2502.19811, 2025{\natexlab{a}}.
\newblock \doi{10.48550/ARXIV.2502.19811}.
\newblock URL \url{https://doi.org/10.48550/arXiv.2502.19811}.

\bibitem[Zhang et~al.(2025{\natexlab{b}})Zhang, Lin, Yao, Hu, Meng, Liu, Men, Yang, Li, Li, Lu, Liu, Chen, Xu, Yu, Wang, Fan, Zhong, Yuan, Zhang, Zhang, Liu, Wang, Fang, He, Liu, Li, Su, Qiu, Pang, Yan, Jiang, Huang, Yin, You, Wei, Wang, Hong, Chen, Chen, Wang, Zheng, Wang, Liu, Dong, Zhang, Pan, Wu, Wu, Guan, Tao, Fu, Xu, Wang, Lai, Wu, Zhou, Yang, and Du]{kimi-linear}
Yu~Zhang, Zongyu Lin, Xingcheng Yao, Jiaxi Hu, Fanqing Meng, Chengyin Liu, Xin Men, Songlin Yang, Zhiyuan Li, Wentao Li, Enzhe Lu, Weizhou Liu, Yanru Chen, Weixin Xu, Longhui Yu, Yejie Wang, Yu~Fan, Longguang Zhong, Enming Yuan, Dehao Zhang, Yizhi Zhang, T.~Y. Liu, Haiming Wang, Shengjun Fang, Weiran He, Shaowei Liu, Yiwei Li, Jianlin Su, Jiezhong Qiu, Bo~Pang, Junjie Yan, Zhejun Jiang, Weixiao Huang, Bohong Yin, Jiacheng You, Chu Wei, Zhengtao Wang, Chao Hong, Yutian Chen, Guanduo Chen, Yucheng Wang, Huabin Zheng, Feng Wang, Yibo Liu, Mengnan Dong, Zheng Zhang, Siyuan Pan, Wenhao Wu, Yuhao Wu, Longyu Guan, Jiawen Tao, Guohong Fu, Xinran Xu, Yuzhi Wang, Guokun Lai, Yuxin Wu, Xinyu Zhou, Zhilin Yang, and Yulun Du.
\newblock Kimi linear: An expressive, efficient attention architecture.
\newblock \emph{CoRR}, abs/2510.26692, 2025{\natexlab{b}}.
\newblock \doi{10.48550/ARXIV.2510.26692}.
\newblock URL \url{https://doi.org/10.48550/arXiv.2510.26692}.

\bibitem[Zhao et~al.(2025)Zhao, Zhou, Zhang, Deng, Xu, Liu, Yu, Li, and Zhao]{deepep}
Chenggang Zhao, Shangyan Zhou, Liyue Zhang, Chengqi Deng, Zhean Xu, Yuxuan Liu, Kuai Yu, Jiashi Li, and Liang Zhao.
\newblock Deepep: an efficient expert-parallel communication library.
\newblock \url{https://github.com/deepseek-ai/DeepEP}, 2025.

\bibitem[Zheng et~al.(2025{\natexlab{a}})Zheng, Bao, Hou, Zheng, Fang, Huang, Li, Duanmu, Chen, Xu, Guo, Zheng, Jiang, Di, Wang, Ye, Lin, Chang, Lu, Liang, Zhai, and Liu]{triton-dist}
Size Zheng, Wenlei Bao, Qi~Hou, Xuegui Zheng, Jin Fang, Chenhui Huang, Tianqi Li, Haojie Duanmu, Renze Chen, Ruifan Xu, Yifan Guo, Ningxin Zheng, Ziheng Jiang, Xinyi Di, Dongyang Wang, Jianxi Ye, Haibin Lin, Li-Wen Chang, Liqiang Lu, Yun Liang, Jidong Zhai, and Xin Liu.
\newblock Triton-distributed: Programming overlapping kernels on distributed ai systems with the triton compiler, 2025{\natexlab{a}}.
\newblock URL \url{https://arxiv.org/abs/2504.19442}.

\bibitem[Zheng et~al.(2025{\natexlab{b}})Zheng, Fang, Zheng, Hou, Bao, Zheng, Jiang, Wang, Ye, Lin, Chang, and Liu]{tilelink}
Size Zheng, Jin Fang, Xuegui Zheng, Qi~Hou, Wenlei Bao, Ningxin Zheng, Ziheng Jiang, Dongyang Wang, Jianxi Ye, Haibin Lin, Li-Wen Chang, and Xin Liu.
\newblock Tilelink: Generating efficient compute-communication overlapping kernels using tile-centric primitives, 2025{\natexlab{b}}.
\newblock URL \url{https://arxiv.org/abs/2503.20313}.

\bibitem[Zhu et~al.(2025)Zhu, Jiang, Jin, Wu, Stuardo, Wang, Zhang, Zhou, Wei, Cheng, et~al.]{megascale-infer}
Ruidong Zhu, Ziheng Jiang, Chao Jin, Peng Wu, Cesar~A Stuardo, Dongyang Wang, Xinlei Zhang, Huaping Zhou, Haoran Wei, Yang Cheng, et~al.
\newblock Megascale-infer: Serving mixture-of-experts at scale with disaggregated expert parallelism.
\newblock \emph{arXiv preprint arXiv:2504.02263}, 2025.

\end{thebibliography}




\end{document}